\documentclass[aps,prl,twocolumn,superscriptaddress,footinbib]{revtex4-2}

\usepackage{graphicx}  
\usepackage{subfigure}
\usepackage{gensymb}
\usepackage{dcolumn} 
\usepackage{bm}
\usepackage{amssymb}   
\usepackage{amsfonts}   
\usepackage{pifont}
\usepackage{comment}
\usepackage[colorlinks,linkcolor=blue,anchorcolor=blue,citecolor=blue,urlcolor=blue]{hyperref}
\usepackage{xcolor}
\usepackage{soul}
\usepackage{tabularx,booktabs,ragged2e,multirow}
\newcolumntype{C}{>{\centering\arraybackslash}X}
\usepackage{multirow}
\usepackage{amsmath}
\usepackage{mathrsfs}
\usepackage{mathcomp}
\usepackage{textcomp}
\usepackage{dsfont}
\usepackage{esint}
\usepackage{braket}
\usepackage{textgreek}
\usepackage{lipsum}
\usepackage{marvosym} 
\usepackage{centernot}
\usepackage{cancel}
\usepackage{dashrule}

\DeclareMathOperator{\Det}{Det}

\DeclareMathOperator{\sgn}{sgn}

\begin{document}
\preprint{APS/123-QED}

\title{In-plane anomalous features in the 3D quantum Hall regime}

\author{Ming Lu}
\affiliation{Beijing Academy of Quantum Information Sciences, Beijing 100193, China}
\author{Xiao-Xiao Zhang}
\email{xxzhang@hust.edu.cn}
\affiliation{Wuhan National High Magnetic Field Center and School of Physics, Huazhong University of Science and Technology, Wuhan 430074, China}


\newcommand{\br}{{\bm r}}
\newcommand{\bk}{{\bm k}}
\newcommand{\bq}{{\bm q}}
\newcommand{\bp}{{\bm p}}
\newcommand{\bv}{{\bm v}}
\newcommand{\bmm}{{\bm m}}
\newcommand{\bA}{{\bm A}}
\newcommand{\bE}{{\bm E}}
\newcommand{\bB}{{\bm B}}
\newcommand{\bH}{{\bm H}}
\newcommand{\bd}{{\bm d}}
\newcommand{\bzero}{{\bm 0}}
\newcommand{\bOmega}{{\bm \Omega}}
\newcommand{\bsigma}{{\bm \sigma}}
\newcommand{\bJ}{{\bm J}}
\newcommand{\bL}{{\bm L}}
\newcommand{\bS}{{\bm S}}
\newcommand{\cA}{{\mathcal A}}
\newcommand{\cB}{{\mathcal B}}
\newcommand{\cC}{{\mathcal C}}
\newcommand\cD{\mathcal{D}}
\newcommand{\cE}{{\mathcal E}}
\newcommand{\cG}{{\mathcal G}}
\newcommand{\cH}{{\mathcal H}}
\newcommand{\cI}{{\mathcal I}}
\newcommand{\cK}{{\mathcal K}}
\newcommand{\cM}{{\mathcal M}}
\newcommand{\cP}{{\mathcal P}}
\newcommand{\cT}{{\mathcal T}}
\newcommand{\cV}{{\mathcal V}}
\newcommand{\cm}{{\mathpzc m}}
\newcommand{\cn}{{\mathpzc n}}
\newcommand\dd{\mathrm{d}}
\newcommand\ii{\mathrm{i}}
\newcommand\ee{\mathrm{e}}
\newcommand\zz{\mathtt{z}}
\makeatletter
\let\newtitle\@title
\let\newauthor\@author
\def\ExtendSymbol#1#2#3#4#5{\ext@arrow 0099{\arrowfill@#1#2#3}{#4}{#5}}
\newcommand\LongEqual[2][]{\ExtendSymbol{=}{=}{=}{#1}{#2}}
\newcommand\LongArrow[2][]{\ExtendSymbol{-}{-}{\rightarrow}{#1}{#2}}
\newcommand{\cev}[1]{\reflectbox{\ensuremath{\vec{\reflectbox{\ensuremath{#1}}}}}}
\newcommand{\red}[1]{\textcolor{red}{#1}} 
\newcommand{\blue}[1]{{\leavevmode\color{blue}#1}} 
\newcommand{\green}[1]{\textcolor{orange}{#1}} 
\newcommand{\mytitle}[1]{\textcolor{orange}{\textit{#1}}}
\newcommand{\mycomment}[1]{} 
\newcommand{\note}[1]{ \textbf{\color{blue}#1}}
\newcommand{\warn}[1]{ \textbf{\color{red}#1}}

\makeatother

\begin{abstract}
Studies of the 3D quantum Hall effect (QHE) have
primarily emphasized transport features that mimic the well-established 2D QHE. In this work, we show that qualitatively new features arise when an in-plane magnetic field is applied to a 3D Weyl semimetal in the quantum Hall regime. An unexpected Hall quantum oscillation, distinct from the Weyl-orbit oscillation, coexists with the QHE, along with an unquantized two-terminal magnetoresistance. Moreover, unconventional antichiral transmission enables a peculiar disorder-robust negative longitudinal resistance. Quantization tunable by the lead configuration is further found in this transport geometry. A unique type of nonlocal quantum backscattering channels underlies these phenomena. Our work demonstrates a breakdown of the topological characterization of transport even with 3D Chern numbers and reveals hidden 3D QHE transport properties. It opens a new class of transport measurements and phenomena.
\end{abstract}

\maketitle

\let\oldaddcontentsline\addcontentsline
\renewcommand{\addcontentsline}[3]{}

\mytitle{Introduction}.--
Magnetotransport and Hall quantization have been of central importance in modern solid-state physics since the discovery of the quantum Hall effect (QHE) in 2D\cite{Klitzing_1980,QHE1990,von_Klitzing_2020}. 
Significant theoretical and experimental efforts have sought to extend the effect to 3D, for instance, in systems with correlated charge or spin order, quantum confinement, or topological semimetallic states\cite{Halperin1987,Montambaux1990,Tang2019,Qin2020,Galeski_2021,Uchida2017,Schumann2018,Lin2019a,Wang2017a,Zhang2018,Nishihaya2019,Nishihaya2021,Zhang2021a}. Among these, the 3D QHE in Weyl semimetals (WSMs), which host linearly dispersing Weyl points (WPs), is of particular interest. It is an intrinsic effect that requires no engineering of the dispersion in the third dimension and combines the QHE topology with the material's native topology\cite{Wang2017a,Li2020a,Chen_2021,Zhang2022,Zhang2018,Nishihaya2021,Nakazawa2024}.
In a WSM under a magnetic field, a 3D semiclassical Weyl orbit can form, consisting of top and bottom surface Fermi arc states connected by chiral Landau levels (CLLs)\cite{Potter2014,Zhang2016b,Moll2016,Zheng2017,Zhang2021a}. These CLLs arise from the bulk WPs and propagate electrons back and forth along the field direction. This mechanism leads to a Weyl-orbit quantum oscillation (QO) when electrons traverse the entire orbit. Deep in the quantum limit, the complete quantization into Fermi-arc Landau levels results in the Hall quantization observed in the geometry of Fig.~\ref{Fig:cartoon}(a) and exemplified in Supplemental Material~\ref{App:background}. A key feature of this 3D QHE is the emergence of a single pair of diagonally counterpropagating hinge-like states\cite{Wang2017a,Li2020a, Zhang2022}.

Thus far, transport studies have focused on this traditional Hall-bar geometry with an out-of-plane field [Fig.~\ref{Fig:cartoon}(a)]. Consequently, the measured Hall quantization scarcely differs from the established 2D QHE\cite{Datta1995}.
However, the extension from two to three dimensions should have more profound physical consequences and introduce transport features unique to 3D systems\cite{Zhang2022, Zhang2025, Zhang2025a}. This is important as WSMs are highly anisotropic due to the presence of WPs and the applied magnetic field $\bB$. 
Given that similar quantum states exist in Dirac semimetals (DSMs), it also remains unclear whether their respective 3D QHEs are distinguishable\cite{Zhang2018,Nishihaya2021}. Moreover, an experimental challenge persists: distinguishing the intrinsic 3D mechanism from various (quasi-)2D effects, such as geometry- or thickness-dependent unconventional QHEs in metals and DSMs, or surface QHEs in systems tunable between WSM and topological insulator phases\cite{Zhang2021a,Cheng2020a,Uchida2017,Lin2019a,Yoshimi2025,belopolski2025}. 
Therefore, there is a strong need to identify truly intrinsic 3D features in the QHE regime that are distinct from, or even go beyond, the established out-of-plane Hall quantization.

Hereafter, we measure resistance $R_{\alpha\beta}$ with current along the $\beta$-direction and voltage drop along the $\alpha$-direction. 
Considering the physical picture for the $zx$-geometry in Fig.~\ref{Fig:cartoon}(a), which is known to yield a quantized $R_{zx}$, should we expect an identically quantized $R_{yx}$ in the $yx$-geometry in Fig.~\ref{Fig:cartoon}(b)? The four transverse voltage leads in the $yx$-geometry appear to be connected to the hinge states, much as in the $zx$-geometry. 
This rarely addressed question points to the system's 3D nature: the transport geometry and lead arrangement become crucial, as these degrees of freedom do not exist in a 2D context. We therefore focus on this unconventional in-plane-field $yx$-geometry, where the Hall-bar plane is parallel to $\bB$ [Fig.~\ref{Fig:cartoon}(b)]. Surprisingly, new 3D features emerge even though no magnetic flux threads through the $yx$-measurement plane. 
We simulate transport in a six-terminal Hall bar, where current $I$ flows along the $x$-axis, using a wavefunction scattering approach\cite{Fisher1981,Groth2014}. 
We refer to the quantized $R_{zx}=(V_1-V_5)/I$ in the $zx$-geometry as the (3D) QHE or $zx$-QHE. $V_i$ is the voltage measured at lead-$i$.
Our focus is on the $yx$-geometry, where the longitudinal, Hall, and two-terminal resistances are respectively defined as $R_{xx}=(V_1-V_2)/I$, $R_{yx}=(V_1-V_5)/I$, and $\tilde{R}_{xx}=(V_0-V_3)/I$, consistent with conventions in QHE measurements \cite{Beenakker1991,Datta1995} (\ref{App:background}).
Specifically, we consider a minimal WSM model
\begin{align}
    H(\bm{k}) &= \sum_{i=x,y,z}2 D_i(1-\cos{k_i})\sigma_0 + A(\sin{k_x}\sigma_x + \sin{k_y}\sigma_y) \nonumber \\
    &+2M[(1-\cos k_\text{w}) - \sum_{i=x,y,z} (1-\cos k_i)]\sigma_z, \label{eq:H_1band1}
\end{align}
with two WPs at $(0, 0, \pm k_\text{w})$ along the $k_z$-axis and under the field $\bm{B}=B\hat{y}$\cite{WeylDiracReview,Nagaosa_2020,Lv2021}. 
The Fermi energy 
is set at the WPs. Onsite disorder is distributed uniformly in $[-W/2, W/2]$.

\begin{figure}[t]
\includegraphics[width=8.6cm]{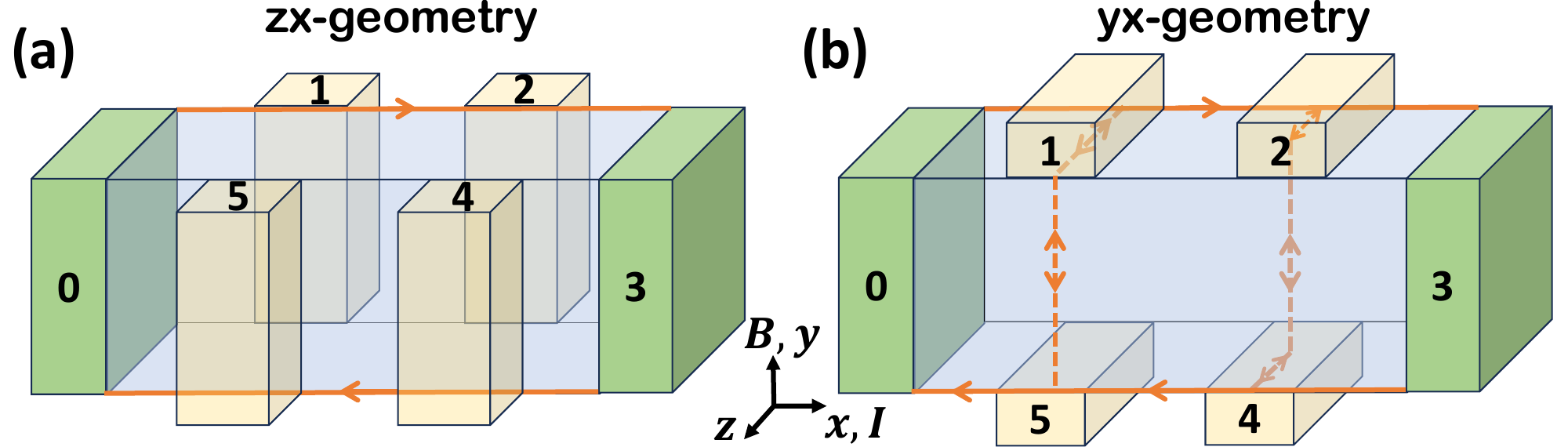}
\caption{Two possible Hall-bar configurations in 3D with the magnetic field (a) perpendicular or (b) parallel to the measurement plane. Current (Voltage) leads are in green (yellow). Solid orange lines denote the top (right-going) and bottom (left-going) hinge states; bidirectional dashed orange lines in (b) represent vertical chiral Landau levels (CLLs) connecting conducting channels near the lead-sample interface. Together, they form a unique type of 3D nonlocal quantum backscattering paths.}\label{Fig:cartoon}
\end{figure}

\begin{figure}[t]
\includegraphics[width=8.7cm]{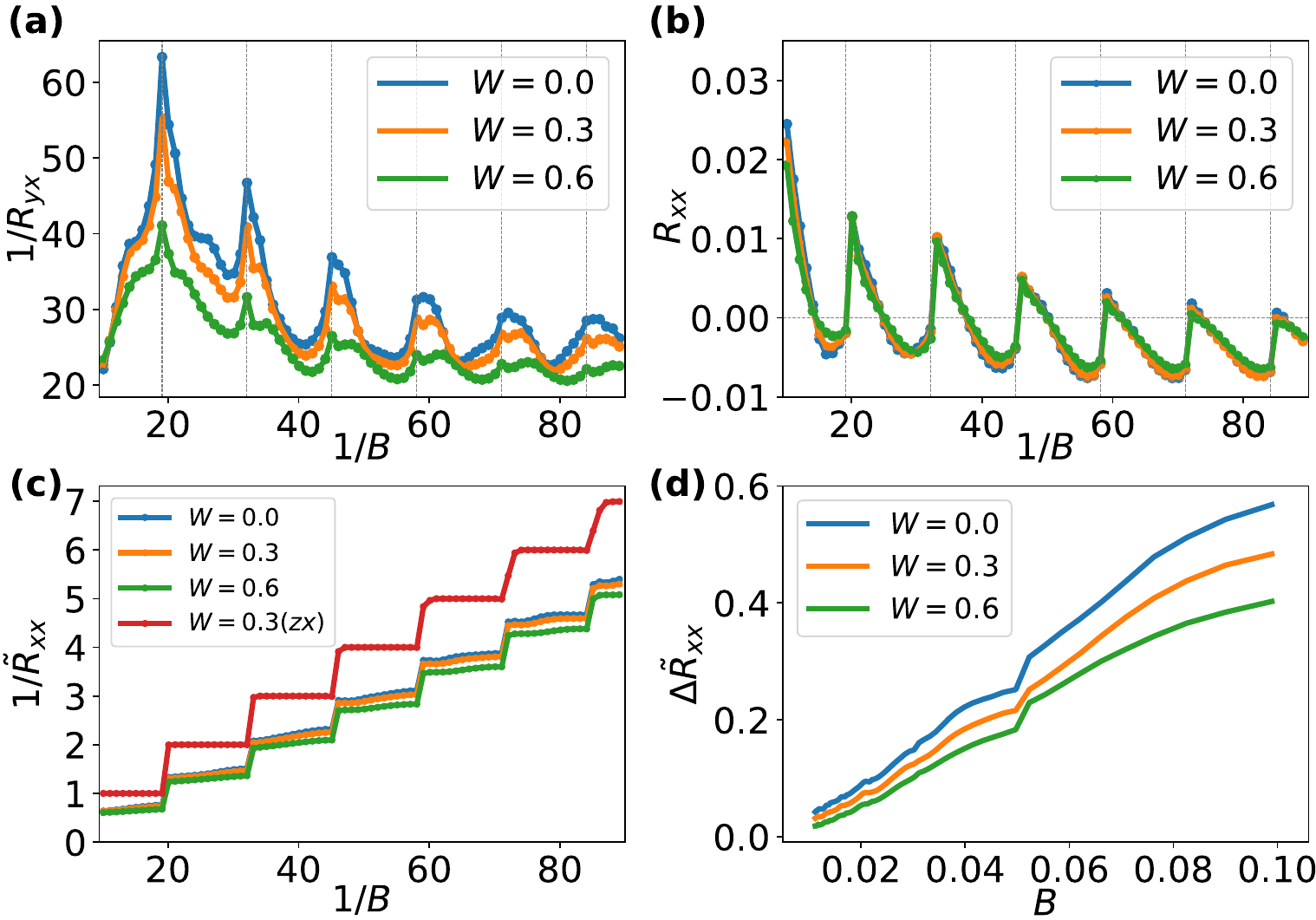}
\caption{(a) The inverse Hall resistance $1/R_{yx}$ and (b) the longitudinal resistance $R_{xx}$ from the $yx$-geometry in Fig.~\ref{Fig:cartoon}(b). The vertical dashed lines mark the magnetic fields of $zx$-QHE plateau transitions from the measurement in Fig.~\ref{Fig:cartoon}(a), indicating that all phenomena exist in the QHE regime of the system. 
(c) Inverse two-terminal resistance $\tilde{R}_{xx}$ in $yx$-geometry; red line: its counterpart in $zx$-geometry shows quantization instead. (d) Residual two-terminal resistance $\Delta\tilde{R}_{xx}$ approximately linear in $B$. System size is $32\times32\times32$ and three disorder strengths $W$ are shown. Model parameters are $A=1, M=0.15, D_x=0.06, D_y=0.03, D_z=0.09, k_\text{w}=\pi/2$.}\label{Fig:Ryx_Rxx}
\end{figure}

\mytitle{QHE-coexistent Hall quantum oscillation}.--
As shown in Fig.~\ref{Fig:Ryx_Rxx}(a,b), and hereafter measured in units of $h/e^2$, both the Hall resistance $R_{yx}$ and longitudinal resistance $R_{xx}$ clearly oscillate with a $1/B$-periodicity. These oscillations coexist with the $zx$-QHE in the same field range for the same system. For reference, Fig.~\ref{Fig:3DQHE} shows the Hall quantization $R_{zx}=1/n$ for $n\in\mathds{N}$, and the corresponding plateau transition fields are marked in Fig.~\ref{Fig:Ryx_Rxx}(a,b). A cyclotron motion, which sustains Hall transport and QOs, is generally not expected here, since the magnetic field is parallel to the $yx$-plane. 
In normal metals with an out-of-plane field, Hall resistance increases with $B$, whereas in the in-plane geometry it changes little. Answering our foregoing question but contradicting these expectations, the in-plane-field $R_{yx}$ is neither trivial nor quantized. Instead, while residing in the QHE regime, it exhibits oscillations atop an overall negative Hall magnetoresistance, evident from the descending peak heights in Fig.~\ref{Fig:Ryx_Rxx}(a).

In the $zx$-geometry, the counterpropagating top and bottom hinge modes are always diagonally separated and are never bridged by quantum states, 
which guarantees quantization and leaves no room for QO.
The 3D nature of the system, however, renders the $yx$-geometry distinct. First, upward- and downward-propagating CLLs, formed from opposite-charge WPs, connect the top and bottom surfaces, effectively ``short-circuiting" the voltage leads and reducing the voltage drop along $y$-direction. Second, $z$-direction propagating modes always exist along the interfaces between the sample and the top/bottom metallic leads, helping the hinge modes bridge their separation along $z$-axis. These two aspects create an intrinsic, coherent, and nonlocal 3D backscattering path, top hinge mode$\to$top lead interface$\to$CLLs$\to$bottom lead interface$\to$bottom hinge mode, as depicted in Fig.~\ref{Fig:cartoon}(b). Consequently, quantization is not expected, as the top and bottom leads and hinge modes are no longer spatially disconnected. 
This Hall QO can be understood as a competition between quantized hinge modes and these short circuits. The number of conducting hinge modes below the Fermi level increases discretely as $B$ decreases, consistent with the $zx$-QHE and visible as jumps in $R_{yx}^{-1}$ across the vertical lines in Fig.~\ref{Fig:Ryx_Rxx}(a). The CLL degeneracy, proportional to $B$, determines the number of parallel short-circuiting channels, contributing a term roughly linear in $B$. Correspondingly, 
the form $R_{zx}^{-1}(B)+aB$, with a fitting parameter $a$, can capture the general trend of $R_{yx}^{-1}$ (see \ref{App:QO}).
These two competing effects lead to the unique QO atop an overall decrease of $R_{yx}$ with increasing $B$. 
Strong disorder weakens the short-circuiting channels, suppressing the oscillation and increasing the resistance.

A persistent issue since the proposal of the 3D QHE has been its relation to the aforementioned Weyl-orbit QO, as both share close physical origins. The short answer is that they typically do not coexist: the semiclassical Weyl-orbit QO becomes less relevant in the full quantum limit of clean samples at high fields, where the arc states quantize into spatially separated hinge states. The present $R_{yx}$-Hall QO thus achieves a rare coexistence by bridging these counterpropagating hinge modes in 3D via metallic interfaces and bulk CLLs. This nonlocal backscattering path is enabled at the cost of destroying quantization in the $yx$-geometry.

\mytitle{Linear two-terminal magnetoresistance}.--
A feature of the 2D QHE is that the two-terminal longitudinal resistance is identical to the Hall resistance, rendering it seemingly useless and often overlooked. Indeed, its counterpart in the $zx$-geometry, $\tilde{R}_{xx}^{(zx)}\equiv R_{zx}$, is perfectly quantized between leads-0,3 according to the number of hinge modes [red line in Fig.~\ref{Fig:Ryx_Rxx}(c)].
This is no longer true for $\tilde{R}_{xx}$ for the $yx$-geometry.
Fig.~\ref{Fig:Ryx_Rxx}(c) shows a clear deviation from quantization, 
which evidently stems from the aforementioned 3D nonlocal backscattering. 
Although the floating transverse leads in a two-terminal measurement are inconsequential in the $zx$-geometry or any 2D QHE, here a 3D-enabled change in measurement geometry completely alters the behavior, revealing the hidden 3D nature of the system.

This unconventional $\tilde{R}_{xx}$, in fact, provides more direct evidence for the backscattering mechanism. 
The inverse Hall resistance $R_{yx}^{-1}$ partly measures the ease of transmission from the top hinge to the bottom one, corresponding to the degree of backscattering and thus bearing a picture similar to that of $\tilde{R}_{xx}$.
Following the previous discussion about $R_{yx}$, we can physically decompose $\tilde{R}_{xx}(B,W)=\alpha(W)\tilde{R}_{xx}^{(zx)}(B)+\Delta\tilde{R}_{xx}$, viewing it as a series connection. Here, $\tilde{R}_{xx}^{(zx)}(B)$, the quantized two-terminal contact resistance, accounts for the stepwise jump.
The residual sample resistance $\Delta\tilde{R}_{xx}$ arises mainly from CLL-assisted backscattering and is also affected by disorder. 
In contrast to the usual quadratic magnetoresistance in metals and semimetals, $\Delta\tilde{R}_{xx}$ exhibits an approximately linear dependence on $B$, as shown in Fig.~\ref{Fig:Ryx_Rxx}(d)\cite{Grosso2014,Zhang2019c,Ali2014}. This signature is expected from the CLL degeneracy, though kinks can appear at plateau transitions due to complex behavior in these regions. 

\mytitle{Negative longitudinal resistance}.--
As in the 2D QHE, the longitudinal resistance of the $zx$-QHE is zero in the plateau regions (Fig.~\ref{Fig:3DQHE}). 
In contrast, the longitudinal resistance $R_{xx}$ in the $yx$-geometry exhibits pronounced QO and even oscillates between positive and negative values [Fig.~\ref{Fig:Ryx_Rxx}(b)]. Finite $R_{xx}$ is now expected due to the backscattering between counterpropagating hinge states. However, $R_{xx}<0$, a \textit{negative static resistance}, is a highly unusual phenomenon, as an intrinsic negative static (rather than differential) resistance is rarely found in quantum materials. Thus, this effect remarkably manifests the potential of 3D quantum phenomena.

To understand the underlying picture, one must recognize that $\bB$ induces a chirality, embodied by the pair of counterpropagating hinge modes in Fig.~\ref{Fig:cartoon}. Let $T_{ij}= T_{i\leftarrow j}$ be the electron transmission probability from lead $j$ to lead $i$. Transmissions such as $T_{10}$ and $T_{20}$ are chiral, as they follow the top hinge mode's propagation, while $T_{13}$ and $T_{23}$ are \textit{antichiral}, opposing the hinge propagation. For simplicity, consider a four-terminal Hall bar [detaching leads 4 and 5 in Fig.~\ref{Fig:cartoon}(b)], which shows a similar sign-oscillating longitudinal resistance. The Landauer-Büttiker formalism gives (\ref{App:negativeR})
\begin{equation}\label{eq:sign}
    \sgn(R_{xx})=\sgn(T_{23}T_{10}-T_{13}T_{20}).
\end{equation}
In the $zx$-QHE [Fig.~\ref{Fig:cartoon}(a)], antichiral transmissions $T_{13}$ and $T_{23}$ are zero within plateaus, guaranteeing $R_{xx}=0$. However, the $yx$-geometry [Fig.~\ref{Fig:cartoon}(b)] enables finite antichiral transmissions $T_{13}$ and $T_{23}$ via the path of bottom hinge-lead interface-CLLs, because the 3D backscattering paths exist for both left- and right-moving incident electrons.
The four-terminal case is special only in that it primarily uses CLLs above the bottom hinge state, as depicted below lead-1 in Fig.~\ref{Fig:cartoon}(b).

Typically, one would expect $T_{23}>T_{13}$ and $T_{10}>T_{20}$, since longer-distance transmission involves more scattering, which gives $R_{xx}>0$ in Eq.~\eqref{eq:sign}. This conclusion assumes the additivity law of transmission probability $\frac{1-T_\mathrm{sum}}{T_\mathrm{sum}}=\frac{1-T_{a}}{T_{a}}+\frac{1-T_{b}}{T_{b}}$, where the total transmission probability $T_\mathrm{sum}$, through two successive scatterers $T_{a}$ and $T_{b}$, is always less than $\mathrm{min}(T_{a}, T_{b})$\cite{Datta2005}. In a bulk normal metal with homogeneous phase-incoherent scatterers, transmission is then dominated by those short paths joining two measuring terminals, because more distant paths experience much more scattering. One can then well approximate transmission $3\rightarrow1$ as a concatenation of $3\rightarrow2$ and $2\rightarrow1$, by viewing $T_{23}>T_{13}$ as $T_{a}>T_\mathrm{sum}$.

This intuition fails, however, in a fully 3D system with highly anisotropic, phase-coherent quantum states.
The $3\rightarrow1$ transmission in Fig.~\ref{Fig:cartoon}(b) is not merely a sequence of $3\rightarrow2\rightarrow1$. 
The nonlocal backscattering path via lead-3$\to$bottom hinge$\to$lead-1 provides an unconventional coherent contribution. The above simple identification of antichiral $T_{23},T_{13}$ as $T_{a},T_\mathrm{sum}$ becomes invalid  
and indeed we find an anomalous relation $T_{23}<T_{13}$. Besides, the chiral transmissions satisfy the normal relation $T_{20}<T_{10}$ as they have no 3D path to reroute. It is this anomalous antichiral transmission that crucially enables $R_{xx}<0$ via Eq.~\eqref{eq:sign}. The six-terminal case can be analyzed in the same spirit.

From a more direct perspective, the 3D nonlocal paths in Fig.~\ref{Fig:cartoon}(b) enable an intriguing top-bottom transmission nonreciprocity, $T_{15}>T_{51}$ and $T_{42}>T_{24}$ (\ref{App:negativeR}). Consequently, lead-1 (lead-2) equilibrates with more (fewer) electrons from the drain lead-3, which are transported by the bottom hinge modes and the CLLs above lead-5 (lead-4). Given that carriers from the drain have a lower chemical potential than the source lead-0 or the top hinge, this imbalance sustains chemical potentials $\mu_1<\mu_2$ at leads 1 and 2, resulting in $R_{xx}<0$.
The CLL degeneracy guarantees the abundance of these 3D paths; the momentum-space WP separation further suppresses scattering between counterpropagating CLLs. These make this phenomenon remarkably robust against disorder, as seen in Fig.~\ref{Fig:Ryx_Rxx}(b).

Furthermore, this negative resistance is found to be absent in the closely related DSM, which has long been deemed equivalent in the 3D QHE (\ref{App:negativeR}). The nonlocal quantum channel proposed here thus enables a significant and sharp distinction between WSMs and DSMs.
Finally, we prove that the two-terminal resistance $\tilde{R}_{xx}$ is positive-definite, thus respecting thermodynamic laws. Physically, its measuring terminals are also the current electrodes. Therefore, the path of the current can be directly associated with two successive scatterers $T_a,T_b$ in the additivity law, corresponding to the contact resistance and the sample resistance prescribed earlier.

\begin{figure}[t]
\includegraphics[width=8.8cm]{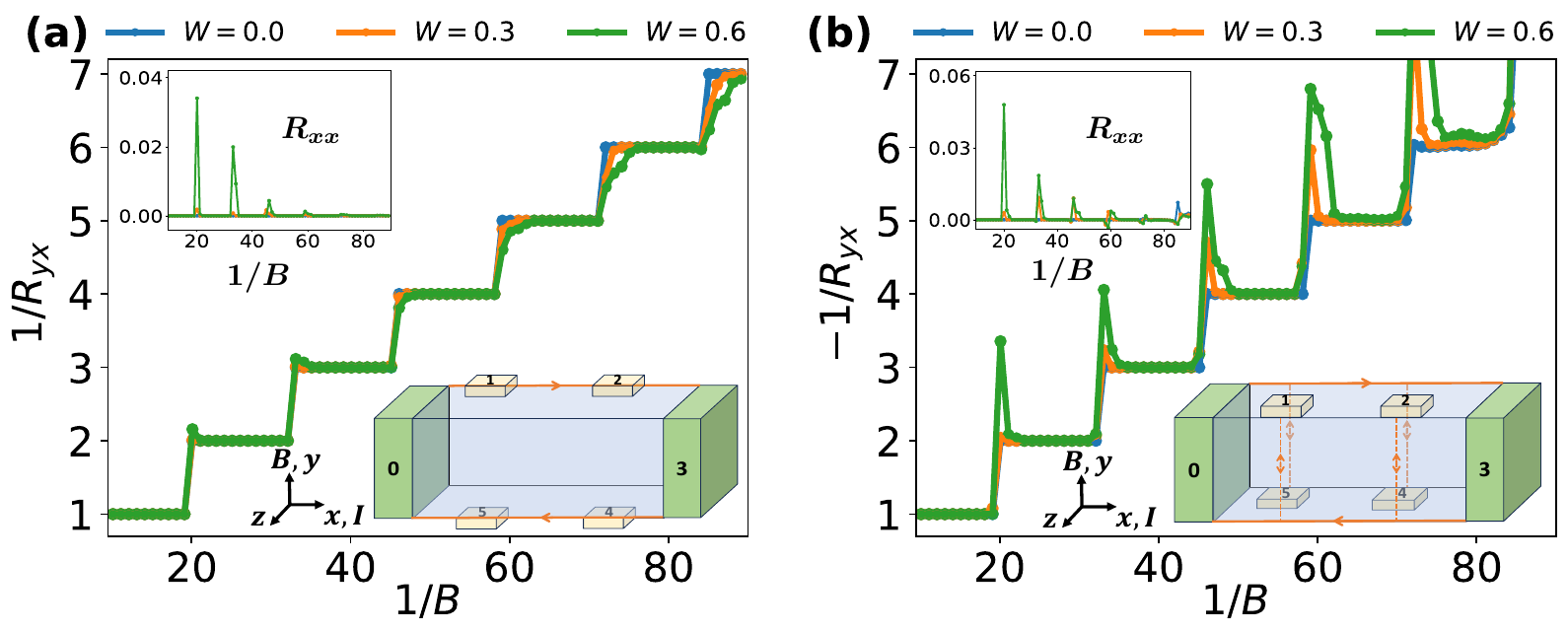}
\caption{The inverse Hall resistance $1/R_{yx}$ for two related narrow-lead configurations in the lower right insets (see Fig.~\ref{Fig:enlarged_insets} for enlarged illustration), in which of (b) CLLs as vertical dashed lines connect voltage leads to hinge modes above or below. 
Corresponding longitudinal resistance $R_{xx}$ is shown in the upper left insets.
}\label{Fig:Rxy_recover_quantized}
\end{figure}

\mytitle{Recovery of quantized Hall effect}.--
The next question is whether quantization can be recovered in this $yx$-geometry. As shown in the lower insets of Fig.~\ref{Fig:Rxy_recover_quantized}(a,b), two simple configurations can disable the 3D nonlocal paths. 
Here, the transverse voltage leads are narrow and do not overlap when projected onto the $zx$-plane. This prevents the formation of complete CLL-assisted short-circuiting or backscattering paths. Indeed, when the voltage leads are adjacent to the hinge states, as in Fig.~\ref{Fig:Rxy_recover_quantized}(a), both a Hall quantization of $R_{yx}$ and a vanishing $R_{xx}$ are restored, as in the conventional $zx$-QHE. 
A similar result, but with a sign-reversed $R_{yx}$, is achieved in the distinct configuration of Fig.~\ref{Fig:Rxy_recover_quantized}(b). Here, the voltage leads are at the far edges away from the hinge states, resulting in a reversed lead-hinge connection via CLLs (\ref{App:ExpConfig}). 
Plenty of CLL channels near the far edges maintain the lead-hinge connection and the robust quantization, though they become more susceptible to disorder near plateau transitions.

Consider a hybrid system that mixes the configurations of Fig.~\ref{Fig:cartoon}(b) and any of Fig.~\ref{Fig:Rxy_recover_quantized} along the $x$-axis, i.e., wide leads on the left and narrow leads on the right [e.g., Fig.~\ref{Fig:hybrid-device}(b)]. For a current in the $\hat{x}$-direction, the left region permits backscattering while the right region does not. This is analogous to a 2D QHE sample where the left region is constricted by a backscattering split gate; the established result predicts that quantization remains intact if measured on the right\cite{Houten1988,Beenakker1991}. 
In our 3D hybrid setup, however, the corresponding quantization of $R_{yx}$ measured on the right fails because the 3D nonlocal paths create an equilibration imbalance in the right region.
Surprisingly, $R_{xx}<0$ persists in this case and is \textit{always} negative for any $B$, strongly enhancing the generality of the negative longitudinal resistance (\ref{App:ExpConfig}).

\newcommand{\blap}[1]{\vbox to 13.5pt{\hbox{#1}\vss}}
\newcommand{\cmark}{\ding{51}}
\newcommand{\xmark}{\ding{55}}
\begin{table}[!hbtp]
    \centering
    \begin{tabular}{c|c|c}
topological characterization & \multicolumn{2}{c}{match/breakdown in transport} \\
 \cmidrule{1-3}
 & \cmark & \xmark \\
         $\cC_x=0$ &  $R_{zy}= 0$ & $R_{yz}\neq0$ \mycomment{small} \\
         $\cC_y=n$ &  $R_{zx}=1/n$  & $R_{xz}$ unquantized  \\
         $\cC_z=\mathrm{AHE}$ &  $R_{xy}=\mathrm{AHE}$ & $R_{yx}=\mathrm{QO}\textrm{ or } 1/n$ \\
    \end{tabular}
    \caption{Topological characterization based on three 3D Chern numbers $\cC_{x,y,z}$ and its comparison with transport features. 
    The \cmark-column shows cases where topological numbers match transport, while the \xmark-column shows breakdowns.}
    \label{tab:3DChern}
\end{table}

\mytitle{Breakdown of topological characterization}.--
We now place the above results in a broader context: the topological characterization of this 3D QHE system. For the 2D QHE in the $xy$-plane, a single Chern number $\cC=n$ is sufficient to determine the Hall resistances $R_{xy}=-R_{yx}=1/n$\cite{ChernTKNN1}. Density-wave-based 3D QHEs, which arise from bulk magnetic band properties, require three Chern numbers $\cC_{x,y,z}$ respectively for transport in the $yz$-, $zx$-, and $xy$-planes\cite{Halperin1987,Montambaux1990}. One might naturally expect the WSM-based 3D QHE to be characterized similarly. Using a real-space Chern-marker method suitable for 3D systems, we evaluate the three 3D Chern numbers, shown in Table~\ref{tab:3DChern} (\ref{App:transportvstopo}). We employ two related formulations for periodic and open boundary conditions 
to handle the field-induced anisotropy\cite{Prodan2009,Prodan2011,Bianco2011,Weise2006,Varjas2020}. Besides vanishing $\cC_x$ and $R_{zy}$, we find an unquantized $\cC_z$ corresponding to the anomalous Hall effect (AHE) in $R_{xy}$, and a quantized $\cC_y$ for the $zx$-QHE in $R_{zx}$. 

However, the correspondence between 3D topological invariants and transport properties breaks down when we switch the current and voltage directions. Besides a nonvanishing $R_{yz}$ dependent on system details, both $R_{xz}$ and other oblique measurement configurations in the $zx$-plane lose their QHE behavior due to inevitable metallic side surfaces\cite{Zhang2022}. Moreover, the in-plane $R_{yx}$ shows no resemblance to the AHE predicted by $\cC_z$. Instead, it exhibits distinct possibilities as we have discussed: the QHE-coexistent QO or the recovered quantization.
Therefore, in Table~\ref{tab:3DChern}, a quantized (unquantized) Chern number can correspond to an unquantized (quantized) Hall response. These features demonstrate an inefficacy of the standard topological characterization of transport, in stark contrast to any 2D-like QHE or the density-wave 3D QHE. This breakdown originates from the 3D anisotropic alignment of the WPs and Fermi arcs (along the $z$-direction) relative to the magnetic field ($B\hat{y}$), which in turn defines the hinge modes (along the $x$-direction). This generates 3D anisotropic quantum states and 3D-enabled nonlocal backscattering paths, making the device geometry and lead configuration nontrivially relevant to the transport outcome.

Our discussion so far has focused on a minimal model where the WP pair is aligned normal to both $\bB$ and the current. However, since the crucial 3D nonlocal path is an intrinsic quantum channel of the system, these conditions can be largely relaxed, justifying the universality of our findings. As noted above, the perfect $zx$-QHE may disappear for an oblique current due to metallic side surfaces. Differently and remarkably, most phenomena proposed here persist with only quantitative modifications, including the peculiar negative longitudinal resistance which might be considered fragile (\ref{App:rotated}). These phenomena originate from hidden 3D physics in the QHE regime but avoid direct observation of the quantization; thus, they demonstrate the merit of exploring transport beyond the established, yet limited, window of Hall quantization.

\mytitle{Summary.}-- 
We have examined transport in a WSM-based 3D QHE system using an unconventional in-plane-field geometry. Various anomalous features are found in the QHE regime, surpassing the simple Hall quantization that is hardly discernible from conventional QHEs. 
They all highlight a new, 3D-only nonlocal backscattering trajectory and ultimately point to a fundamental breakdown of standard topological characterization. The results establish a novel class of 3D quantum transport highly relevant to current experimental challenges.


\begin{acknowledgments}
We appreciate the helpful discussions with Zhi-Qiang Zhang and Hailong Li. The research was supported by the National Natural Science Foundation of China (Grant No.~12574172 and Grant No.~12204044). The computation was partially completed in the HPC Platform of Huazhong University of Science and Technology.
\end{acknowledgments}\mycomment{\Yinyang}


\bibliography{reference.bib}  

@Article{ChernTKNN1,
  author           = {Thouless, D. J. and Kohmoto, M. and Nightingale, M. P. and den Nijs, M.},
  journal          = {Phys. Rev. Lett.},
  title            = {Quantized {Hall} Conductance in a Two-Dimensional Periodic Potential},
  year             = {1982},
  month            = {Aug},
  pages            = {405--408},
  volume           = {49},
  doi              = {10.1103/PhysRevLett.49.405},
  issue            = {6},
  modificationdate = {2022-01-24T19:52:27},
  numpages         = {0},
  publisher        = {American Physical Society},
  url              = {https://link.aps.org/doi/10.1103/PhysRevLett.49.405},
}

@Article{WeylDiracReview,
  author           = {{Armitage}, N.~P. and {Mele}, E.~J. and {Vishwanath}, A.},
  journal          = {Rev. Mod. Phys.},
  title            = {{Weyl and {Dirac} Semimetals in Three Dimensional Solids}},
  year             = {2018},
  month            = may,
  number           = {1},
  pages            = {015001},
  volume           = {90},
  doi              = {10.1103/RevModPhys.90.015001},
  modificationdate = {2022-01-24T19:54:59},
  publisher        = {American Physical Society ({APS})},
}

@Book{Datta2005,
  author    = {Supriyo Datta},
  publisher = {Cambridge University Press},
  title     = {Quantum Transport: Atom to Transistor},
  year      = {2005},
  address   = {Cambridge},
  doi       = {10.1017/cbo9781139164313},
}

@Article{Weise2006,
  author    = {Alexander Wei{\ss}e and Gerhard Wellein and Andreas Alvermann and Holger Fehske},
  journal   = {Reviews of Modern Physics},
  title     = {The kernel polynomial method},
  year      = {2006},
  month     = {mar},
  number    = {1},
  pages     = {275--306},
  volume    = {78},
  doi       = {10.1103/revmodphys.78.275},
  publisher = {American Physical Society ({APS})},
}

@Article{Zhang2021a,
  author           = {Cheng Zhang and Yi Zhang and Hai-Zhou Lu and X. C. Xie and Faxian Xiu},
  journal          = {Nature Reviews Physics},
  title            = {Cycling Fermi arc electrons with {Weyl} orbits},
  year             = {2021},
  month            = {jul},
  number           = {9},
  pages            = {660--670},
  volume           = {3},
  doi              = {10.1038/s42254-021-00344-z},
  modificationdate = {2022-01-24T19:54:28},
  publisher        = {Springer Science and Business Media {LLC}},
}

@Article{Wang2017a,
  author           = {C.{\hspace{0.167em}}M. Wang and Hai-Peng Sun and Hai-Zhou Lu and X.{\hspace{0.167em}}C. Xie},
  journal          = {Physical Review Letters},
  title            = {{3D} Quantum {Hall} Effect of Fermi Arcs in Topological Semimetals},
  year             = {2017},
  month            = {sep},
  number           = {13},
  pages            = {136806},
  volume           = {119},
  doi              = {10.1103/physrevlett.119.136806},
  modificationdate = {2022-01-24T19:52:27},
  publisher        = {American Physical Society ({APS})},
}

@Article{Li2020a,
  author           = {Hailong Li and Haiwen Liu and Hua Jiang and X.{\hspace{0.167em}}C. Xie},
  journal          = {Physical Review Letters},
  title            = {{3D} Quantum {Hall} Effect and a Global Picture of Edge States in {Weyl} Semimetals},
  year             = {2020},
  month            = {jul},
  number           = {3},
  pages            = {036602},
  volume           = {125},
  doi              = {10.1103/physrevlett.125.036602},
  modificationdate = {2022-01-24T19:54:28},
  publisher        = {American Physical Society ({APS})},
}

@Book{Datta1995,
  author    = {Supriyo Datta},
  publisher = {Cambridge University Press},
  title     = {Electronic Transport in Mesoscopic Systems},
  year      = {1995},
  address   = {Cambridge},
  month     = {sep},
  doi       = {10.1017/cbo9780511805776},
}

@Article{Groth2014,
  author    = {Christoph W Groth and Michael Wimmer and Anton R Akhmerov and Xavier Waintal},
  journal   = {New Journal of Physics},
  title     = {Kwant: a software package for quantum transport},
  year      = {2014},
  month     = {jun},
  number    = {6},
  pages     = {063065},
  volume    = {16},
  doi       = {10.1088/1367-2630/16/6/063065},
  publisher = {{IOP} Publishing},
}

@Article{Halperin1987,
  author    = {Bertrand I. Halperin},
  journal   = {Japanese Journal of Applied Physics},
  title     = {Possible States for a Three-Dimensional Electron Gas in a Strong Magnetic Field},
  year      = {1987},
  month     = {jan},
  number    = {S3-3},
  pages     = {1913},
  volume    = {26},
  doi       = {10.7567/jjaps.26s3.1913},
  publisher = {{IOP} Publishing},
}

@Article{Montambaux1990,
  author           = {G. Montambaux and M. Kohmoto},
  journal          = {Physical Review B},
  title            = {Quantized {Hall} effect in three dimensions},
  year             = {1990},
  month            = {jun},
  number           = {16},
  pages            = {11417--11421},
  volume           = {41},
  doi              = {10.1103/physrevb.41.11417},
  modificationdate = {2022-01-24T19:52:27},
  publisher        = {American Physical Society ({APS})},
}

@Article{Tang2019,
  author           = {Fangdong Tang and Yafei Ren and Peipei Wang and Ruidan Zhong and John Schneeloch and Shengyuan A. Yang and Kun Yang and Patrick A. Lee and Genda Gu and Zhenhua Qiao and Liyuan Zhang},
  journal          = {Nature},
  title            = {Three-dimensional quantum {Hall} effect and metal{\textendash}insulator transition in {ZrTe{$_5$}}},
  year             = {2019},
  month            = {may},
  number           = {7757},
  pages            = {537--541},
  volume           = {569},
  doi              = {10.1038/s41586-019-1180-9},
  modificationdate = {2022-01-24T19:50:39},
  publisher        = {Springer Science and Business Media {LLC}},
}

@Article{Zhang2018,
  author           = {Cheng Zhang and Yi Zhang and Xiang Yuan and Shiheng Lu and Jinglei Zhang and Awadhesh Narayan and Yanwen Liu and Huiqin Zhang and Zhuoliang Ni and Ran Liu and Eun Sang Choi and Alexey Suslov and Stefano Sanvito and Li Pi and Hai-Zhou Lu and Andrew C. Potter and Faxian Xiu},
  journal          = {Nature},
  title            = {Quantum {Hall} effect based on {Weyl} orbits in {Cd{$_3$}As{$_2$}}},
  year             = {2018},
  month            = {dec},
  number           = {7739},
  pages            = {331--336},
  volume           = {565},
  doi              = {10.1038/s41586-018-0798-3},
  modificationdate = {2022-01-24T19:52:27},
  publisher        = {Springer Science and Business Media {LLC}},
}

@Article{Potter2014,
  author           = {Andrew C. Potter and Itamar Kimchi and Ashvin Vishwanath},
  journal          = {Nature Communications},
  title            = {Quantum oscillations from surface Fermi arcs in {Weyl} and {Dirac} semimetals},
  year             = {2014},
  month            = {oct},
  number           = {1},
  pages            = {5161},
  volume           = {5},
  doi              = {10.1038/ncomms6161},
  modificationdate = {2022-01-24T19:54:59},
  publisher        = {Springer Science and Business Media {LLC}},
}

@Article{Zhang2016b,
  author           = {Yi Zhang and Daniel Bulmash and Pavan Hosur and Andrew C. Potter and Ashvin Vishwanath},
  journal          = {Scientific Reports},
  title            = {Quantum oscillations from generic surface Fermi arcs and bulk chiral modes in {Weyl} semimetals},
  year             = {2016},
  month            = {apr},
  number           = {1},
  pages            = {23741},
  volume           = {6},
  doi              = {10.1038/srep23741},
  modificationdate = {2022-01-24T19:54:28},
  publisher        = {Springer Science and Business Media {LLC}},
}

@Article{Lin2019a,
  author           = {Ben-Chuan Lin and Shuo Wang and Steffen Wiedmann and Jian-Ming Lu and Wen-Zhuang Zheng and Dapeng Yu and Zhi-Min Liao},
  journal          = {Phys. Rev. Lett.},
  title            = {Observation of an Odd-Integer Quantum {Hall} Effect from Topological Surface States in {Cd{$_3$}As{$_2$}}},
  year             = {2019},
  month            = {jan},
  number           = {3},
  pages            = {036602},
  volume           = {122},
  doi              = {10.1103/physrevlett.122.036602},
  modificationdate = {2022-01-24T19:48:49},
  publisher        = {American Physical Society ({APS})},
}

@Article{Qin2020,
  author           = {Fang Qin and Shuai Li and Z.{\hspace{0.167em}}Z. Du and C.{\hspace{0.167em}}M. Wang and Wenqing Zhang and Dapeng Yu and Hai-Zhou Lu and X.{\hspace{0.167em}}C. Xie},
  journal          = {Phys. Rev. Lett.},
  title            = {Theory for the Charge-Density-Wave Mechanism of {3D} Quantum {Hall} Effect},
  year             = {2020},
  month            = {nov},
  number           = {20},
  pages            = {206601},
  volume           = {125},
  doi              = {10.1103/physrevlett.125.206601},
  modificationdate = {2022-01-24T19:51:06},
  publisher        = {American Physical Society ({APS})},
}

@Article{Lv2021,
  author           = {B.{\hspace{0.167em}}Q. Lv and T. Qian and H. Ding},
  journal          = {Reviews of Modern Physics},
  title            = {Experimental perspective on three-dimensional topological semimetals},
  year             = {2021},
  month            = {apr},
  number           = {2},
  pages            = {025002},
  volume           = {93},
  creationdate     = {2021-11-17T09:41:44},
  doi              = {10.1103/revmodphys.93.025002},
  modificationdate = {2021-11-17T09:43:09},
  publisher        = {American Physical Society ({APS})},
}

@Article{Fisher1981,
  author           = {Daniel S. Fisher and Patrick A. Lee},
  journal          = {Physical Review B},
  title            = {Relation between conductivity and transmission matrix},
  year             = {1981},
  month            = {jun},
  number           = {12},
  pages            = {6851--6854},
  volume           = {23},
  creationdate     = {2021-11-29T14:16:44},
  doi              = {10.1103/physrevb.23.6851},
  modificationdate = {2021-11-29T14:18:14},
  publisher        = {American Physical Society ({APS})},
}

@Article{Nishihaya2021,
  author           = {Shinichi Nishihaya and Masaki Uchida and Yusuke Nakazawa and Markus Kriener and Yasujiro Taguchi and Masashi Kawasaki},
  journal          = {Nature Communications},
  title            = {Intrinsic coupling between spatially-separated surface Fermi-arcs in {Weyl} orbit quantum {Hall} states},
  year             = {2021},
  month            = {may},
  number           = {1},
  pages            = {2572},
  volume           = {12},
  creationdate     = {2021-12-02T16:07:53},
  doi              = {10.1038/s41467-021-22904-8},
  modificationdate = {2022-01-24T19:54:28},
  publisher        = {Springer Science and Business Media {LLC}},
}

@Article{Schumann2018,
  author           = {Timo Schumann and Luca Galletti and David{\hspace{0.167em}}A. Kealhofer and Honggyu Kim and Manik Goyal and Susanne Stemmer},
  journal          = {Physical Review Letters},
  title            = {Observation of the Quantum {Hall} Effect in Confined Films of the Three-Dimensional {Dirac} Semimetal {Cd{$_3$}As{$_2$}}},
  year             = {2018},
  month            = {jan},
  number           = {1},
  pages            = {016801},
  volume           = {120},
  creationdate     = {2021-12-02T17:12:50},
  doi              = {10.1103/physrevlett.120.016801},
  modificationdate = {2022-01-24T19:52:27},
  publisher        = {American Physical Society ({APS})},
}

@Article{Uchida2017,
  author           = {Masaki Uchida and Yusuke Nakazawa and Shinichi Nishihaya and Kazuto Akiba and Markus Kriener and Yusuke Kozuka and Atsushi Miyake and Yasujiro Taguchi and Masashi Tokunaga and Naoto Nagaosa and Yoshinori Tokura and Masashi Kawasaki},
  journal          = {Nature Communications},
  title            = {Quantum {Hall} states observed in thin films of {Dirac} semimetal {Cd{$_3$}As{$_2$}}},
  year             = {2017},
  month            = {dec},
  number           = {1},
  pages            = {2274},
  volume           = {8},
  creationdate     = {2021-12-02T21:45:16},
  doi              = {10.1038/s41467-017-02423-1},
  modificationdate = {2022-01-24T19:52:27},
  publisher        = {Springer Science and Business Media {LLC}},
}

@Book{QHE1990,
  editor           = {Richard E. Prange and Steven M. Girvin},
  publisher        = {Springer New York},
  title            = {The Quantum {Hall} Effect},
  year             = {1990},
  address          = {New York},
  edition          = {2nd},
  creationdate     = {2022-01-23T11:10:35},
  doi              = {10.1007/978-1-4612-3350-3},
  modificationdate = {2022-01-24T19:52:27},
}

@Article{von_Klitzing_2020,
  author           = {Klaus von Klitzing and Tapash Chakraborty and Philip Kim and Vidya Madhavan and Xi Dai and James McIver and Yoshinori Tokura and Lucile Savary and Daria Smirnova and Ana Maria Rey and Claudia Felser and Johannes Gooth and Xiaoliang Qi},
  journal          = {Nature Reviews Physics},
  title            = {40 years of the quantum {Hall} effect},
  year             = {2020},
  month            = {jul},
  number           = {8},
  pages            = {397--401},
  volume           = {2},
  creationdate     = {2022-01-23T11:14:34},
  doi              = {10.1038/s42254-020-0209-1},
  modificationdate = {2022-01-24T19:52:27},
  publisher        = {Springer Science and Business Media {LLC}},
}

@Article{Galeski_2021,
  author           = {S. Galeski and T. Ehmcke and R. Wawrzy{\'{n}}czak and P. M. Lozano and K. Cho and A. Sharma and S. Das and F. Küster and P. Sessi and M. Brando and R. Küchler and A. Markou and M. König and P. Swekis and C. Felser and Y. Sassa and Q. Li and G. Gu and M. V. Zimmermann and O. Ivashko and D. I. Gorbunov and S. Zherlitsyn and T. Förster and S. S. P. Parkin and J. Wosnitza and T. Meng and J. Gooth},
  journal          = {Nature Communications},
  title            = {Origin of the quasi-quantized {Hall} effect in {ZrTe{$_5$}}},
  year             = {2021},
  month            = {may},
  number           = {1},
  pages            = {3197},
  volume           = {12},
  creationdate     = {2022-01-23T11:54:29},
  doi              = {10.1038/s41467-021-23435-y},
  modificationdate = {2022-01-24T19:52:27},
  publisher        = {Springer Science and Business Media {LLC}},
}

@Article{Klitzing_1980,
  author           = {K. v. Klitzing and G. Dorda and M. Pepper},
  journal          = {Physical Review Letters},
  title            = {New Method for High-Accuracy Determination of the Fine-Structure Constant Based on Quantized {Hall} Resistance},
  year             = {1980},
  month            = {aug},
  number           = {6},
  pages            = {494--497},
  volume           = {45},
  creationdate     = {2022-01-23T11:59:06},
  doi              = {10.1103/physrevlett.45.494},
  modificationdate = {2022-01-24T19:52:27},
  publisher        = {American Physical Society ({APS})},
}

@Article{Chen_2021,
  author           = {Rui Chen and Tianyu Liu and C.{\hspace{0.167em}}M. Wang and Hai-Zhou Lu and X.{\hspace{0.167em}}C. Xie},
  journal          = {Physical Review Letters},
  title            = {Field-Tunable One-Sided Higher-Order Topological Hinge States in {Dirac} Semimetals},
  year             = {2021},
  month            = {aug},
  number           = {6},
  pages            = {066801},
  volume           = {127},
  creationdate     = {2022-03-01T15:28:50},
  doi              = {10.1103/physrevlett.127.066801},
  modificationdate = {2022-03-01T15:30:56},
  publisher        = {American Physical Society ({APS})},
}

@Article{Nagaosa_2020,
  author           = {Naoto Nagaosa and Takahiro Morimoto and Yoshinori Tokura},
  journal          = {Nature Reviews Materials},
  title            = {Transport, magnetic and optical properties of {Weyl} materials},
  year             = {2020},
  month            = {jun},
  number           = {8},
  pages            = {621--636},
  volume           = {5},
  creationdate     = {2022-03-05T13:13:30},
  doi              = {10.1038/s41578-020-0208-y},
  modificationdate = {2022-03-05T13:29:51},
  publisher        = {Springer Science and Business Media {LLC}},
}

@Article{Zhang2022,
  author    = {Xiao-Xiao Zhang and Naoto Nagaosa},
  journal   = {Nano Letters},
  title     = {Anisotropic Three-Dimensional Quantum Hall Effect and Magnetotransport in Mesoscopic Weyl Semimetals},
  year      = {2022},
  month     = {mar},
  number    = {7},
  pages     = {3033--3039},
  volume    = {22},
  doi       = {10.1021/acs.nanolett.2c00296},
  publisher = {American Chemical Society ({ACS})},
}

@Article{Zhang2019c,
  author           = {Zhang, ShengNan and Wu, QuanSheng and Liu, Yi and Yazyev, Oleg V.},
  journal          = {Physical Review B},
  title            = {Magnetoresistance from Fermi surface topology},
  year             = {2019},
  issn             = {2469-9969},
  month            = jan,
  number           = {3},
  pages            = {035142},
  volume           = {99},
  creationdate     = {2024-09-18T14:35:41},
  doi              = {10.1103/physrevb.99.035142},
  modificationdate = {2024-09-18T14:35:41},
  publisher        = {American Physical Society (APS)},
}

@Article{Ali2014,
  author           = {Ali, Mazhar N. and Xiong, Jun and Flynn, Steven and Tao, Jing and Gibson, Quinn D. and Schoop, Leslie M. and Liang, Tian and Haldolaarachchige, Neel and Hirschberger, Max and Ong, N. P. and Cava, R. J.},
  journal          = {Nature},
  title            = {Large, non-saturating magnetoresistance in WTe2},
  year             = {2014},
  issn             = {1476-4687},
  month            = sep,
  number           = {7521},
  pages            = {205--208},
  volume           = {514},
  creationdate     = {2024-09-18T14:35:55},
  doi              = {10.1038/nature13763},
  modificationdate = {2024-09-18T14:35:55},
  publisher        = {Springer Science and Business Media LLC},
}

@Book{Grosso2014,
  author           = {Giuseppe Grosso and Giuseppe Pastori Parravicini},
  publisher        = {Elsevier},
  title            = {Solid state physics},
  year             = {2014},
  address          = {Waltham, Massachusetts},
  edition          = {2},
  isbn             = {9780123850300},
  creationdate     = {2024-09-18T15:12:08},
  doi              = {10.1016/C2010-0-66724-1},
  modificationdate = {2024-09-18T15:16:03},
  pagetotal        = {1873},
  ppn_gvk          = {1659306647},
}

@Article{Cheng2020a,
  author    = {Cheng, Shu-guang and Jiang, Hua and Sun, Qing-Feng and Xie, X. C.},
  journal   = {Physical Review B},
  title     = {Quantum Hall effect in wedge-shaped samples},
  year      = {2020},
  issn      = {2469-9969},
  month     = aug,
  number    = {7},
  pages     = {075304},
  volume    = {102},
  doi       = {10.1103/physrevb.102.075304},
  publisher = {American Physical Society (APS)},
}

@Article{Varjas2020,
  author    = {Varjas, Dániel and Fruchart, Michel and Akhmerov, Anton R. and Perez-Piskunow, Pablo M.},
  journal   = {Physical Review Research},
  title     = {Computation of topological phase diagram of disordered <mml:math xmlns:mml="http://www.w3.org/1998/Math/MathML"><mml:mrow><mml:msub><mml:mi>Pb</mml:mi><mml:mrow><mml:mn>1</mml:mn><mml:mo>−</mml:mo><mml:mi>x</mml:mi></mml:mrow></mml:msub><mml:msub><mml:mi>Sn</mml:mi><mml:mi>x</mml:mi></mml:msub><mml:mi>Te</mml:mi></mml:mrow></mml:math> using the kernel polynomial method},
  year      = {2020},
  issn      = {2643-1564},
  month     = feb,
  number    = {1},
  pages     = {013229},
  volume    = {2},
  doi       = {10.1103/physrevresearch.2.013229},
  publisher = {American Physical Society (APS)},
}

@Article{Prodan2009,
  author    = {Prodan, Emil},
  journal   = {Physical Review B},
  title     = {Robustness of the spin-Chern number},
  year      = {2009},
  issn      = {1550-235X},
  month     = sep,
  number    = {12},
  pages     = {125327},
  volume    = {80},
  doi       = {10.1103/physrevb.80.125327},
  publisher = {American Physical Society (APS)},
}

@Article{Prodan2011,
  author    = {Prodan, Emil},
  journal   = {Journal of Physics A: Mathematical and Theoretical},
  title     = {Disordered topological insulators: a non-commutative geometry perspective},
  year      = {2011},
  issn      = {1751-8121},
  month     = feb,
  number    = {11},
  pages     = {113001},
  volume    = {44},
  doi       = {10.1088/1751-8113/44/11/113001},
  publisher = {IOP Publishing},
}

@Article{Bianco2011,
  author    = {Bianco, Raffaello and Resta, Raffaele},
  journal   = {Physical Review B},
  title     = {Mapping topological order in coordinate space},
  year      = {2011},
  issn      = {1550-235X},
  month     = dec,
  number    = {24},
  pages     = {241106},
  volume    = {84},
  doi       = {10.1103/physrevb.84.241106},
  publisher = {American Physical Society (APS)},
}

@InBook{Beenakker1991,
  author           = {Beenakker, C.W.J. and van Houten, H.},
  pages            = {1--228},
  publisher        = {Elsevier},
  title            = {Quantum Transport in Semiconductor Nanostructures},
  year             = {1991},
  booktitle        = {Semiconductor Heterostructures and Nanostructures},
  creationdate     = {2024-10-18T09:59:54},
  doi              = {10.1016/s0081-1947(08)60091-0},
  issn             = {0081-1947},
  modificationdate = {2024-10-18T09:59:54},
}

@Article{Houten1988,
  author    = {van Houten, H. and Beenakker, C. W. J. and van Loosdrecht, P. H. M. and Thornton, T. J. and Ahmed, H. and Pepper, M. and Foxon, C. T. and Harris, J. J.},
  journal   = {Physical Review B},
  title     = {Four-terminal magnetoresistance of a two-dimensional electron-gas constriction in the ballistic regime},
  year      = {1988},
  issn      = {0163-1829},
  month     = may,
  number    = {14},
  pages     = {8534--8536},
  volume    = {37},
  doi       = {10.1103/physrevb.37.8534},
  publisher = {American Physical Society (APS)},
}

@Article{Nakazawa2024,
  author    = {Nakazawa, Yusuke and Kurihara, Ryosuke and Miyazawa, Masatoshi and Nishihaya, Shinichi and Kriener, Markus and Tokunaga, Masashi and Kawasaki, Masashi and Uchida, Masaki},
  journal   = {Journal of the Physical Society of Japan},
  title     = {Edge and Bulk States in Weyl-Orbit Quantum Hall Effect as Studied by Corbino Measurements},
  year      = {2024},
  issn      = {1347-4073},
  month     = feb,
  number    = {2},
  pages     = {023706},
  volume    = {93},
  doi       = {10.7566/jpsj.93.023706},
  publisher = {Physical Society of Japan},
}

@Article{Moll2016,
  author    = {Moll, Philip J. W. and Nair, Nityan L. and Helm, Toni and Potter, Andrew C. and Kimchi, Itamar and Vishwanath, Ashvin and Analytis, James G.},
  journal   = {Nature},
  title     = {Transport evidence for Fermi-arc-mediated chirality transfer in the Dirac semimetal Cd3As2},
  year      = {2016},
  issn      = {1476-4687},
  month     = jul,
  number    = {7611},
  pages     = {266--270},
  volume    = {535},
  doi       = {10.1038/nature18276},
  publisher = {Springer Science and Business Media LLC},
}

@Article{Zheng2017,
  author    = {Zheng, Guolin and Wu, Min and Zhang, Hongwei and Chu, Weiwei and Gao, Wenshuai and Lu, Jianwei and Han, Yuyan and Yang, Jiyong and Du, Haifeng and Ning, Wei and Zhang, Yuheng and Tian, Mingliang},
  journal   = {Physical Review B},
  title     = {Recognition of Fermi-arc states through the magnetoresistance quantum oscillations in Dirac semimetal Cd3As2 nanoplates},
  year      = {2017},
  issn      = {2469-9969},
  month     = sep,
  number    = {12},
  pages     = {121407},
  volume    = {96},
  doi       = {10.1103/physrevb.96.121407},
  publisher = {American Physical Society (APS)},
}

@Article{Nishihaya2019,
  author           = {Nishihaya, Shinichi and Uchida, Masaki and Nakazawa, Yusuke and Kurihara, Ryosuke and Akiba, Kazuto and Kriener, Markus and Miyake, Atsushi and Taguchi, Yasujiro and Tokunaga, Masashi and Kawasaki, Masashi},
  journal          = {Nature Communications},
  title            = {Quantized surface transport in topological Dirac semimetal films},
  year             = {2019},
  issn             = {2041-1723},
  month            = jun,
  number           = {1},
  pages            = {2564},
  volume           = {10},
  creationdate     = {2024-11-05T10:34:45},
  doi              = {10.1038/s41467-019-10499-0},
  modificationdate = {2024-11-05T10:43:32},
  publisher        = {Springer Science and Business Media LLC},
}

@Article{Belopolski2025,
  author    = {Belopolski, Ilya and Watanabe, Ryota and Sato, Yuki and Yoshimi, Ryutaro and Kawamura, Minoru and Nagahama, Soma and Zhao, Yilin and Shao, Sen and Jin, Yuanjun and Kato, Yoshihiro and Okamura, Yoshihiro and Zhang, Xiao-Xiao and Fujishiro, Yukako and Takahashi, Youtarou and Hirschberger, Max and Tsukazaki, Atsushi and Takahashi, Kei S. and Chiu, Ching-Kai and Chang, Guoqing and Kawasaki, Masashi and Nagaosa, Naoto and Tokura, Yoshinori},
  journal   = {Nature},
  title     = {Synthesis of a semimetallic Weyl ferromagnet with point Fermi surface},
  year      = {2025},
  issn      = {1476-4687},
  month     = jan,
  number    = {8048},
  pages     = {1078--1083},
  volume    = {637},
  doi       = {10.1038/s41586-024-08330-y},
  publisher = {Springer Science and Business Media LLC},
}

@Article{Yoshimi2025,
  author    = {Yoshimi, Ryutaro and Kurihara, Ryosuke and Okamura, Yoshihiro and Handa, Hikaru and Ogawa, Naoki and Kawamura, Minoru and Tsukazaki, Atsushi and Takahashi, Kei S. and Kawasaki, Masashi and Takahashi, Youtarou and Tokunaga, Masashi and Tokura, Yoshinori},
  journal   = {Physical Review Letters},
  title     = {Emergence of Ferroelectric Topological Insulator as Verified by Quantum Hall Effect of Surface States in (Sn,Pb,In)Te Films},
  year      = {2025},
  issn      = {1079-7114},
  month     = apr,
  number    = {17},
  pages     = {176602},
  volume    = {134},
  doi       = {10.1103/physrevlett.134.176602},
  publisher = {American Physical Society (APS)},
}

@Article{Zhang2025,
  author    = {Zhang, Zhi-Qiang and Li, Yu-Hang and Lu, Ming and Liu, Hongfang and Li, Hailong and Jiang, Hua and Xie, X.C.},
  journal   = {Science Bulletin},
  title     = {Three-dimensional quantum anomalous Hall effect in Weyl semimetals},
  year      = {2025},
  issn      = {2095-9273},
  month     = sep,
  doi       = {10.1016/j.scib.2025.09.037},
  publisher = {Elsevier BV},
}

@Article{Zhang2025a,
  author        = {Zhang, Zhi-Qiang and Cheng, Shu-Guang and Liu, Hongfang and Li, Hailong and Jiang, Hua and Xie, X. C.},
  journal       = {arXiv},
  title         = {Chern Vector Protected Three-dimensional Quantized Hall Effect},
  year          = {2025},
  month         = {jan},
  archiveprefix = {arXiv},
  copyright     = {arXiv.org perpetual, non-exclusive license},
  doi           = {10.48550/ARXIV.2501.14493},
  eprint        = {2501.14493},
  primaryclass  = {cond-mat.mes-hall},
  publisher     = {arXiv},
}
\let\addcontentsline\oldaddcontentsline

\newpage
\onecolumngrid
\newpage
{
	\center \bf \large 
	Supplemental Material \\
	\large for ``\newtitle"\vspace*{0.1cm}\\ 
	\vspace*{0.5cm}
}
\begin{center}
	Ming Lu$^{1}$ and Xiao-Xiao Zhang$^{2}$\\
	\vspace*{0.15cm}
    \small{$^1$\textit{Beijing Academy of Quantum Information Sciences, Beijing 100193, China}}\\
    \small{$^2$\textit{Wuhan National High Magnetic Field Center and School of Physics, Huazhong University of Science and Technology, Wuhan 430074, China}}\\
	\vspace*{0.25cm}	
\end{center}


\tableofcontents


\setcounter{section}{0}
\setcounter{equation}{0}
\setcounter{figure}{0}
\setcounter{table}{0}
\setcounter{page}{1}
\renewcommand{\theequation}{S\arabic{equation}}
\renewcommand{\thefigure}{S\arabic{figure}}
\renewcommand{\thetable}{S\arabic{table}}
\renewcommand{\theHtable}{Supplement.\thetable}
\renewcommand{\theHfigure}{Supplement.\thefigure}
\renewcommand{\thesection}{SM~\Roman{section}}
\renewcommand{\bibnumfmt}[1]{[S#1]}
\renewcommand{\citenumfont}[1]{S#1}


\section{Measurement convention and 3D QHE}\label{App:background}
\subsection{Hall-bar measurement of QHE}

\begin{figure}[hbt]
\includegraphics[width=7.4cm]{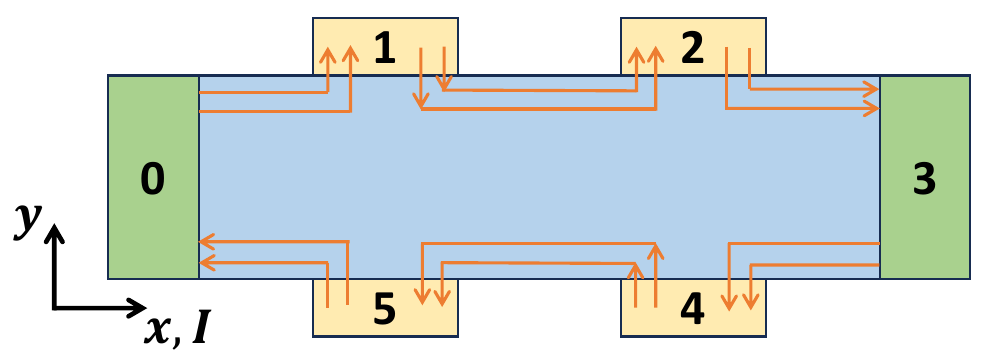}
\caption{A standard six-terminal Hall-bar measurement in a 2D QHE system with two chiral edge states represented by orange arrow lines. The current leads are in green while the voltage leads are in yellow. The lead labelling in Fig.~\ref{Fig:cartoon} follows the same convention herein.}
\label{Fig:2DQH_standard}
\end{figure}

In this section, we clarify the measurement convention used in this study and its relation to the measurement convention in conventional QHE. In Fig.~\ref{Fig:2DQH_standard}, we schematically show the standard six-terminal Hall-bar measurement for a 2D system, e.g., a conventional QHE system, in the $xy$-plane. The net current $I$ flows along $x$-direction. Experimentally, the most interested measurement coefficients include the Hall resistance $R_{yx}$, the longitudinal resistance $R_{xx}$ and the two-terminal resistance $\tilde{R}_{xx}$. They are defined as follows
\begin{equation}\label{eq:standard_R_definition}
    R_{yx} = \frac{V_2-V_4}{I} = \frac{V_1-V_5}{I}, \quad R_{xx} = \frac{V_1-V_2}{I} = \frac{V_5-V_4}{I}, \quad  \tilde{R}_{xx} = \frac{V_0-V_3}{I},
\end{equation}
where $V_i$ is the voltage measured at lead-$i$.

In any plateau regime of a QHE, while the longitudinal resistance vanishes, the Hall resistance $R_{yx}$ and the two terminal resistance $\tilde{R}_{xx}$, effectively measuring the contact resistance, take an identical quantized value\cite{Beenakker1991,Datta1995}. Namely,
\begin{equation}\label{eq:2DQHE}
    R_{yx} = \tilde{R}_{xx}=\frac{1}{n} \frac{h}{e^2}, \quad R_{xx}=0.
\end{equation}

For the six-terminal Hall-bar measurement on a 3D system with magnetic field $B\hat{y}$, given the current direction along $x$, two possible directions exist to attach the voltage leads. If the voltage leads are along $z$-direction, one gets the $zx$-geometry as illustrated in Fig.~\ref{Fig:cartoon}(a); if the voltage leads are along $y$-direction, one gets the $yx$-geometry as illustrated in Fig.~\ref{Fig:cartoon}(b). When projected onto the $zx$- and $yx$-planes, respectively, they are the same as the standard Hall-bar geometry in 2D shown in Fig.~\ref{Fig:2DQH_standard}. Likewise, the definitions for the Hall resistance, the two-terminal resistance, and the longitudinal resistance are unaltered, given by the same form in Eq.~\eqref{eq:standard_R_definition}.

\subsection{3D QHE in WSMs}
\begin{figure}[hbt]
\includegraphics[width=16.8cm]{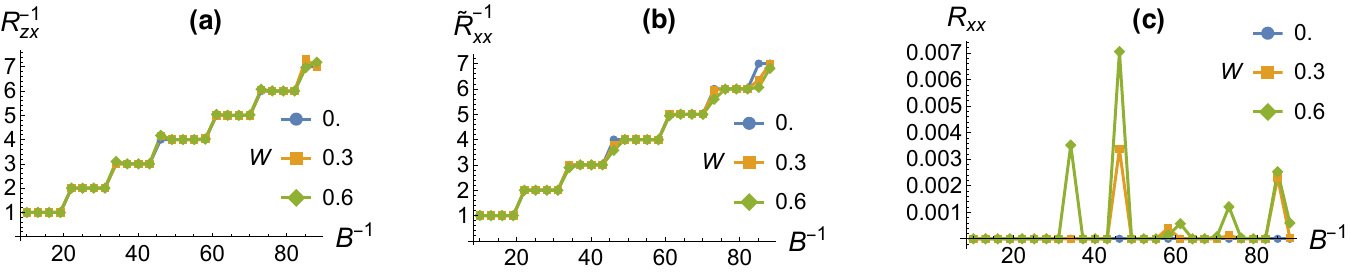}
\caption{Transport feature of the WSM-based 3D QHE measured in the $zx$-geometry in Fig.~\ref{Fig:cartoon}(a). (a) Hall quantization of $R_{zx}$; (b) quantizied two-terminal resistance $\tilde{R}_{xx}$ identical to the Hall resistance $R_{zx}$; (c) vanishing longitudinal resistance $R_{xx}$. Three different disorder strengths are exemplified to show the robustness of the QHE.
}\label{Fig:3DQHE}
\end{figure}

Here, we recapitulate the main features of the 3D QHE in a WSM system, measured in the $zx$-geometry in Fig.~\ref{Fig:cartoon}(a). The Hall resistance, two-terminal resistance, and the longitudinal resistance are successively shown in Fig.~\ref{Fig:3DQHE}(a,b,c). In Fig.~\ref{Fig:3DQHE}(a), perfect Hall quantization is clearly realized and very robust even against strong disorders. In Fig.~\ref{Fig:3DQHE}(b), one observes that the two-terminal resistance almost exactly follows the quantization behavior of Fig.~\ref{Fig:3DQHE}(a). Also, Fig.~\ref{Fig:3DQHE}(c) shows the vanishing longitudinal resistance except for possible deviations at the plateau transitions. 

These features are indeed expected from Eq.~\eqref{eq:2DQHE}, i.e., the 3D QHE, measured in the out-of-plane-field $zx$-geometry in Fig.~\ref{Fig:cartoon}(a), is identical to the 2D conventional QHE in transport phenomena. The only difference is that the 2D $xy$-plane in Fig.~\ref{Fig:2DQH_standard} is replaced by the $xz$-plane in the 3D case in Fig.~\ref{Fig:cartoon}(a). 

The one pair of hinge states in this 3D QHE, illustrated in Fig.~\ref{Fig:cartoon} is a defining feature that is fundamentally different from the conventional 2D QHE. In the 2D case, as illustrated in Fig.~\ref{Fig:2DQH_standard}, there are propagating modes at all (i.e., both) edges along the current direction. 
One might wonder why the 3D case has only two instead of all four possible hinges along the current direction. The physical picture for this seemingly peculiar behavior in 3D is the following: the surface state, say, on the top, is an open Fermi arc and can contribute only half of the cyclotron orbit; more qualitatively, the missing hinge state on the top reflects the inversion symmetry breaking at the surface, which is also the cause of the open Fermi arc. In the conventional 2D QHE, the cyclotron orbit is entirely of a 2D bulk nature and is itself 2D and hence the edge states exist along all the edges.

Although this hinge-state feature can be theoretically verified by calculating the wavefunction distribution\cite{Li2020a,Zhang2022}, it is not readily accessible to experiments. On the other hand, as shown in Fig.~\ref{Fig:3DQHE}, its transport features look identical to the 2D QHE.  
This, in a way, becomes a deficiency since the 3D nature of this state is masked and experimental verification and distinguishing from other (quasi-)2D mechanisms will face obstacles, as mentioned in the main text. This will, however, not be the case in the in-plane-field $yx$-geometry in Fig.~\ref{Fig:cartoon}(b), which is the main focus of the present study.
To facilitate the comparison and correspondence between the out-of-plane-field $zx$-geometry QHE and the in-plane-field $yx$-geometry anomalous features, we mark the QHE plateau transition fields also in Fig.~\ref{Fig:Ryx_Rxx}(a,b).

\section{Approximate characterization of in-plane QO}\label{App:QO}

\begin{figure}[hbt]
\includegraphics[width=5.4cm]{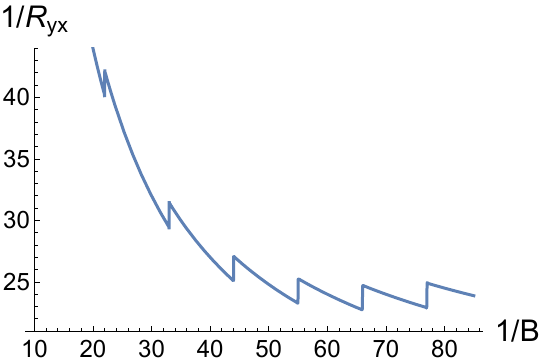}
\caption{Approximate characterization of Fig.~\ref{Fig:Ryx_Rxx}(a) with parameters $c=2,a_1=11,a_2=420$.
}\label{Fig:QOfit}
\end{figure}

The $yx$-geometry QO coexisting with the $zx$-geometry QHE in Fig.~\ref{Fig:Ryx_Rxx}(a) can be approximately captured by a combination of two effects discussed in the main text, the quantized number of hinge modes and the CLL short-circuiting, in the form $R_{yx}^{-1}(B)=R_{zx}^{-1}(B)+aB$. More concretely, we use the integral floor function $\lfloor x \rfloor$ in the following form $R_{yx}^{-1}=c(\lfloor \frac{B^{-1}}{a_1} \rfloor+a_2B)$ to obtain Fig.~\ref{Fig:QOfit} that resembles the overall variation trend in Fig.~\ref{Fig:Ryx_Rxx}(a).

\section{Negative resistance}\label{App:negativeR}

\subsection{Four-terminal Hall bar}
\begin{figure}[hbt]
\includegraphics[width=16.4cm]{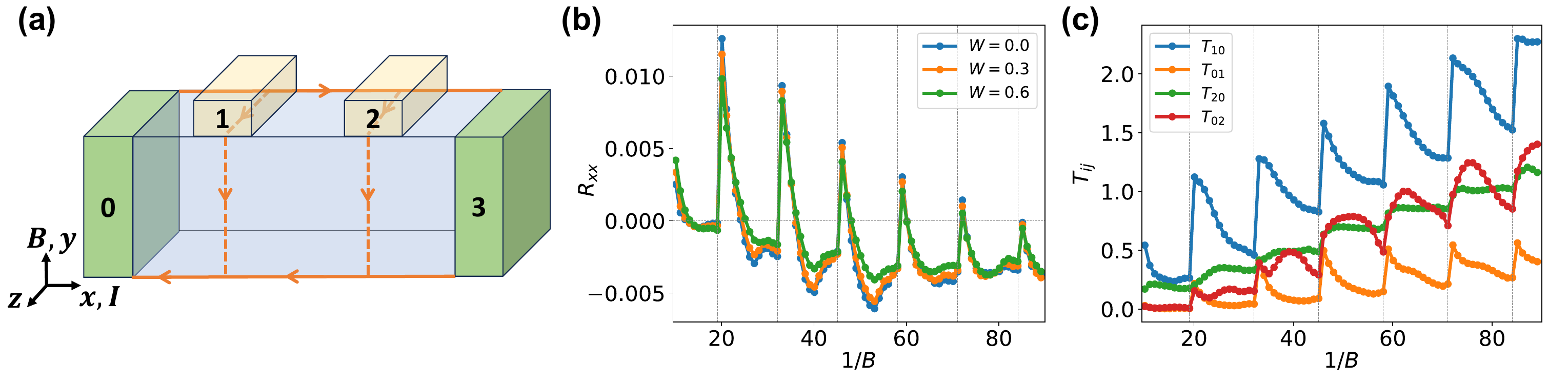}
\caption{(a) Schematics of the four-terminal Hall bar in $yx$-geometry. (b) The longitudinal resistance of $R_{xx}$ with different disorders strength. (c) The transmission coefficients of $T_{13}$ and $T_{23}$. The dashed vertical lines mark the Hall plateau transitions, same as the those in Fig.~\ref{Fig:Ryx_Rxx}(b) in the main text.
}\label{Fig:4terminal}
\end{figure}

According to the Landauer-B\"uttiker formalism, one finds in the four-terminal Hall bar, illustrated in Fig.~\ref{Fig:4terminal}(a), that $I=GV$ with the conductance matrix given by
\begin{equation}
    G=\frac{e^2}{h}\begin{bmatrix}
T_0 & -T_{01} & -T_{02} \\
-T_{10} & T_1 & -T_{12}\\
-T_{20} & -T_{21} & T_2
\end{bmatrix}
\end{equation}
and the current vector $I=(1,0,0)^\mathrm{T}$ as we set the voltage $V_3$ at terminal-3 to zero. $T_i=\tilde\sum_{j=0}^{3}T_{ij}$ ($\tilde\sum$ denoting $j=i$ excluded in the sum) is the total transmission probability inward terminal $i$. In this 3D QHE system, $T_0$ is always quantized as per the number of hinge modes.
Solving the voltage vector $V=(V_0,V_1,V_2)^\mathrm{T}$, one finds the longitudinal resistance
\begin{equation}
    R_{xx} =V_1-V_2= (\frac{e^2}{h})^2\frac{T_{10}T_{23}-T_{20}T_{13}}{\Det[G]}
\end{equation}
where 
\begin{equation}
\begin{split}
    \Det[G](h/e^2)^3 &=T_0T_{21}T_{12}+T_1T_{02}T_{20}+T_2T_{01}T_{10}+T_{21}T_{10}T_{02}+T_{01}T_{12}T_{20}-T_0T_1T_2 \\
    &= T_{02} (T_{13} T_{21}+T_{23} T_1)+T_{01} (T_{12} T_{23}+T_{13} T_2)+T_{03} (T_{12} (T_{20}+T_{23})+(T_{10} +T_{13}) T_2)\\
    &>0
\end{split}
\end{equation}
always holds since all $T$'s are positive. Thus, one arrives at 
\begin{equation}\label{eq:sgn_4T_Rxx}
    \sgn(R_{xx})=\sgn(T_{10}T_{23}-T_{20}T_{13})
\end{equation}
as given in the main text, which can be either positive or negative.

From symmetry considerations similar to that in Sec.~\ref{App:symmetry}, we have $T_{23}=T_{01}$ and $T_{13}=T_{02}$, leading to $\sgn(R_{xx})=\sgn(T_{10}T_{01}-T_{20}T_{02})$. As shown in Fig.~\ref{Fig:4terminal}(c), we can see apparent jumps for the transmission coefficients when the Hall plateau jumps, less notably for the coefficients of $T_{20}$ and $T_{02}$. This results from the fact that the number of chiral edge modes increases by one as the magnetic field crosses a vertical line. Between two adjacent vertical lines, i.e., within one plateau region of the 3D QHE as shown in Fig.~\ref{Fig:Ryx_Rxx}(c), the number of chiral modes is a constant. However, as the magnetic field decreases within one plateau region, the localization of chiral edge modes decreases since the magnetic length increases. This decrease in the field-induced wavefunction localization or confinement is mostly along $y$-axis for the hinge modes\cite{Li2020a,Zhang2022}; therefore, the tunnelling probability from the hinge modes into lead-1 decreases as the spatial weight overlap shrinks, leading to the decrease of $T_{10}=T_{1\leftarrow 0}$. Besides, the decreasing of the magnetic field will make the degeneracy of the vertical chiral Landau levels decrease, decreasing the number of backscattering channels, as schematically shown with the orange arrow lines in Fig.~\ref{Fig:4terminal}(a). This in turn will make the antichiral transmission probability $T_{01}$ decrease as well. 

On the other hand, as mentioned above, fewer electrons from lead-0 will tunnel into lead-1 when the magnetic field decreases, bringing more electrons near lead-2. However, due to the decreasing magnetic confinement, a smaller fraction of these electrons will tunnel into lead 2. When these two contrasting effects are combined, we have a more or less flat, even slightly increasing transmission probability $T_{20}$ within one plateau region. The transmission probability from lead 2 to lead 0 is also affected by two factors. On one side, the decreasing magnetic field will make fewer electrons from lead-2 tunnel into the left moving chiral hinge mode, because of the decreasing of the vertical CLLs; on the other side, fewer vertical CLLs also means fewer electrons from the chiral edge modes tunnelling into lead-1, giving room for more electrons tunnelling directly into lead-0. These two contrasting effects result in a somewhat complicated behavior of $T_{02}$, which increases at first and then decreases within one plateau. All in all, when decreasing the magnetic field within one plateau region, the apparently rapid decrease of $T_{10}T_{01}$ versus the nonmonotonic but less varying $T_{20}T_{02}$ enables negative resistance $R_{xx}$ in a significant portion of each plateau region, as shown in Fig.~\ref{Fig:4terminal}(b).

\subsection{Six-terminal Hall bar}\label{App:6T}
\begin{figure}[hbt]
\includegraphics[width=16.4cm]{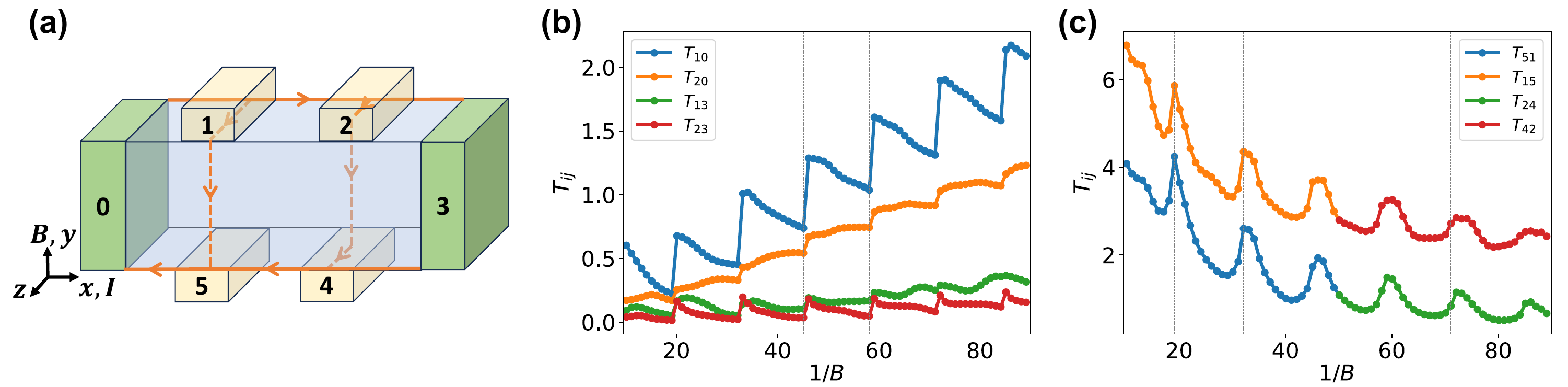}
\caption{(a) Schematics of the six-terminal Hall bar in $yx$-geometry. (b) Transmission coefficients $T_{10}$, $T_{20}$, $T_{13}$ and $T_{23}$.  (c) Transmission coefficients $T_{51}$, $T_{15}$, $T_{24}$, and $T_{42}$. By symmetry considerations in Sec. \ref{App:symmetry}, $T_{51}=T_{24}$ and $T_{15}=T_{42}$ hold. The dashed vertical lines in (b) and (c) mark the Hall plateau transitions, same as the those in Fig.~\ref{Fig:Ryx_Rxx}(b) in the main text.
}\label{Fig:6T-Tij}
\end{figure}

Now we study the more general and complex six-terminal Hall bar case illustrated in Fig.~\ref{Fig:cartoon}(b) and reproduced in Fig.~\ref{Fig:6T-Tij}(a). One finds $I=GV$ with the conductance matrix
\begin{equation}
    G=\frac{e^2}{h}\begin{bmatrix}
T_0 & -T_{01} & -T_{02} & -T_{03} & -T_{04} \\
-T_{10} & T_1 & -T_{12} & -T_{13} & -T_{14}\\
-T_{20} & -T_{21} & T_2 & -T_{23} & -T_{24} \\
-T_{30} & -T_{31} & -T_{32} & T_3 & -T_{34} \\
-T_{40} & -T_{41} & -T_{42} & -T_{43} & T_4 
\end{bmatrix},
\end{equation}
the current vector $I=(1,0,0,-1,0)^\mathrm{T}$ as we set the voltage $V_5$ at terminal-5 to zero, and $T_i=\tilde\sum_{j=0}^{5}T_{ij}$. In this 3D QHE system, $T_0=T_3$ is always quantized as per the number of hinge modes.
Solving the voltage vector $V=(V_0,V_1,V_2,V_3,V_4)^\mathrm{T}$, one finds the longitudinal resistance
\begin{equation}\label{eq:6T_Rxx}
    R_{xx} =V_5-V_4= \frac{h}{e^2}\frac{E_1+E_2}{F}
\end{equation}
where 
\begin{equation}\label{eq:E12}
    E_1=T_{23}(T_{10}+A_1)-T_{13}(T_{20}+A_2),\quad E_2=T_{10}A_2-T_{20}A_1
\end{equation}
with $A_1=T_{14}+T_{15},A_2=T_{14}+T_{51}$ and
\begin{equation}
\begin{split}
    F&=T_{01}^2 (T_{02} + T_{12} + T_{14}) + T_{04}^2 (T_{05} + T_{12} + T_{14}) + 
 T_{02}^2 (T_{01} + T_{14} + T_{21}) + T_{05}^2 (T_{04} + T_{14} + T_{21}) \\
 +& 
 T_{04} T_{05} (T_{01} + T_{02} + T_{03} + T_{12} + 4 T_{14} + T_{15} + T_{21} + T_{24}) \\
 +& T_{01} X + T_{02} [X + T_{01} (T_{12} + 4 T_{14} + T_{15} + T_{21} + T_{24})] \\
 +& T_{03} [2X + T_{01} (T_{02} + T_{12} 
 + 2 T_{14} + T_{15}) + T_{02} (2 T_{14} + T_{21} + T_{24})] \\
 +& T_{04} [X + T_{03} (T_{02} + T_{12} + 2 T_{14} + T_{15}) + T_{01} (T_{02} + 2 (T_{14} + T_{15}))] \\
 +& T_{05} [X + T_{01} (T_{02} + T_{03}) + 2 T_{02} (T_{14} + T_{24}) + T_{03} (2 T_{14} + T_{21} + T_{24})]
\end{split}
\end{equation}
with $X=2 T_{14}^2 + T_{15} T_{21} + T_{12} (T_{14} + T_{24}) + T_{14} (T_{15} + T_{21} + T_{24})$. 
To reach this result for the complex six-terminal case, one needs to take several symmetries of the system setting into account, which is detailed in Sec.~\ref{App:symmetry}. 

Since apparently $F>0$ for all positive $T$'s in Eq.~\eqref{eq:6T_Rxx}, we immediately have 
\begin{equation}\label{eq:sgn_6T_Rxx}
    \sgn(R_{xx})=\sgn(E_1+E_2),
\end{equation}
which, similar to Eq.~\eqref{eq:sgn_4T_Rxx}, can be either positive or negative as per Eq.~\eqref{eq:E12}. We follow a similar approach to see how it can be negative, as in the main text that focuses on nonzero antichiral transmission probabilities. 
Then the discussion is reduced to the following two aspects respectively for $E_1$ and $E_2$ in Eq.~\eqref{eq:E12}.
\begin{itemize}
    \item Firstly, as shown in Fig.~\ref{Fig:6T-Tij}(b), one finds in the calculation that the anomalous relation $T_{23}<T_{13}$ for antichiral transmissions generically holds while the normal $T_{10}>T_{20}$ for chiral transmissions always remains true. This is indeed the same as the four-terminal negative resistance discussed in the main text and thus displays the similarity in terms of the essential physics between the two system settings. Previewing the relation $T_{51}<T_{15}$ discussed right below, one immediately finds that the anomalous $T_{23}<T_{13}$ is again crucial since it is apparently the only factor that drives $E_1<0$ to realize. 

    \item Secondly, as shown in Fig.~\ref{Fig:6T-Tij}(c), one finds in the calculation a second anomalous relation $T_{51}<T_{15}$ always true (as well as $T_{24}<T_{42}$ related by Eq.~\eqref{eq:symmetry_offdiag}), which is apparently the only factor that drives $E_2<0$ to realize. Note that this relation complies with the symmetry analysis around Eq.~\eqref{eq:T51}. Reminiscent of the reasoning for $T_{23}<T_{13}$, this can again be understood from the crucial field- and 3D-enabled large backscattering paths. 
    
    Now in Fig.~\ref{Fig:6T-Tij}(a), both transmissions $5\rightarrow1$ and $1\rightarrow5$ can happen vertically through CLLs above lead-5. Electrons from lead-1 (lead-5) can also go to the current lead-3 (lead-0) directly via the right-going (left-going) hinge mode; in the absence of leads-2,4, such current lead-going transmission probabilities physically do not differ much although leads-1,5 are located leftward; this is because the electron movement is along ballistic channels without extra scattering mechanisms, although this might become less robust if the system were a normal metal with a roughly uniform disorder distribution.  
    However, in the presence of leads-2,4, part of the electrons potentially for the transmission $1\rightarrow5$ will be distracted by the backscattering path connected to leads-2,4 and go into these leads. This univocally reduces $T_{51}$ but does not affect the transmission $5\rightarrow1$ due to the configuration of the backscattering path fixed by the field and the WSM Hamiltonian. 
\end{itemize}
Therefore, in summary, we have both $E_1<0$ and $E_2<0$ enabled and closely related to the antichiral transmissions and iconic 3D backscattering paths, generating $R_{xx}<0$ in this six-terminal case. 

In the main text, we also give an alternative and intuitive explanation of $R_{xx}<0$ for the six-terminal case, which is based on the chemical potentials $\mu_1$ and $\mu_2$ measured by leads-1,2 and modified by the 3D backscattering path. Here, we further comment on its validity and usage.
\begin{itemize}
    \item Firstly, the argument seemingly directly relies on $T_{51}<T_{15}$ and $T_{24}<T_{42}$, which apparently traces mainly to the second point above for $\sgn(E_2)$. However, one should be aware that undoubtedly both the above two points for $E_1$ and $E_2$ play a role. The one for $\sgn(E_1)$ also contributes to the building of the imbalance $T_{51}<T_{15},T_{24}<T_{42}$, because they are, in fact, inseparable and indispensable aspects of the same 3D backscattering paths, i.e., electrons through such paths simultaneously contribute to both aspects. 
    
    \item Secondly, the chemical potential argument can be readily generalized to the four-terminal case. The minor difference is twofold: i) the relevant 3D backscattering paths now mainly involve CLLs above the bottom hinge states; ii) although leads-4,5 are absent, the similar 3D paths described in i) still give rise to the imbalance in the number of drain potential-carrying electrons. 

    \item Lastly, one might wonder if the chemical potential argument is consistent with the sign oscillation of $R_{xx}$ that also shows positive values. In fact, comparing $\Delta T=T_{51}-T_{15}$ and $R_{xx}$ as a function of $B$, one observes that the valley of $\Delta T$ always match the peaks of $R_{xx}$, clearly indicating that the least anomalous imbalance corresponds to the most normal effects upon the resistance. Given that this effect is not the only contributing factor in the transport process, naturally one can have normal $R_{xx}>0$ when it is weak during the oscillation.

    To elaborate on this point, it is interesting to note that $R_{xx}$ in hybrid configurations in Fig.~\ref{Fig:hybrid-device} is almost always negative, compared to the sign-switching behavior in the unhybridized six-terminal case in Fig.~\ref{Fig:Ryx_Rxx}(b). Given the similarity between four-terminal and six-terminal cases as we numerously mentioned, let us take the hybridized six-terminal case in Fig.~\ref{Fig:hybrid-device}(b) as an example below. 
    
    The argument for that hybrid configuration in Sec.~\ref{App:hybrid} is mainly based on one particular 3D nonlocal path i) $\textrm{bottom hinge}\rightarrow1\rightarrow2$ incident from the right, which is simple and clear. The key difference in the unhybridized six-terminal case lies in that another path ii) $\textrm{top hinge}\rightarrow4\rightarrow5$ incident from the left is also functioning. Then there can be the complexity that $\mu_s$-carrying top-hinge electrons also travel to the bottom lead-4 and may eventually reach back towards lead-1; given the foregoing $T_{15}>T_{51}$ that transports more $\mu_d$-carrying electrons to lead-1, there are thus two competing trends. Although we do not expect the effect due to path ii) always being strong as it transports electrons through a longer and almost closed path, it still physically can occasionally outrun the effect due to path i). Obviously, a very similar competing situation takes place for lead-2 as well. Therefore, as we see in the calculation, one by no means expects that $\mu_1<\mu_2$ or $R_{xx}<0$ always holds in the unhybridized six-terminal case. So is the four-terminal case as it only differs in the spatial position of contributing CLLs.

\end{itemize}

\subsection{Positive two-terminal resistance}
In this part, we prove the positive definiteness of the two-terminal resistance $\tilde{R}_{xx}$, regardless of the measurement setting. 
We follow the convention in the foregoing discussion of $R_{xx}$. 
For the four-terminal case, we note that the two-terminal resistance can be cast into the form
\begin{equation}
    \tilde{R}_{xx} = V_0-V_3 = (\frac{e^2}{h})^2\frac{T_{12} (T_{20} + T_{23}) + T_2(T_{10} + T_{13})}{\Det[G]}
\end{equation}
which is readily seen to be always positive.
For the six-terminal case, it can be cast into the form
\begin{equation}
\begin{split}
    \tilde{R}_{xx} =& V_0-V_3 \\
    = &\frac{h}{e^2}\frac{1}{F}\left[T_{02} (T_{04} + 2 T_{14} + T_{21} + T_{24}) + T_{04} (T_{05} + T_{12} + 2 T_{14} + T_{15}) + T_{05} (2 T_{14} + T_{21} + T_{24}) \right.\\
    +& \left.
 2 ( T_{15} T_{21} + T_{12} T_{24}) + 2T_{14} (T_{12}+2 T_{14}+T_{15} + T_{21} + T_{24})+T_{01} (T_{02} + T_{05} + T_{12} + 2 T_{14} + T_{15})\right]
\end{split}
\end{equation}
which is again readily seen to be always positive.

In the main text, we have talked about the decomposition of $\tilde{R}_{xx}$ in a way similar to that of $R_{yx}^{-1}$, shown in \ref{App:QO}. Note that the decomposition is more accurate for $\tilde{R}_{xx}$, exemplified in Fig.~\ref{Fig:Ryx_Rxx}(c,d), because it most directly measures the backscattering, i.e., the difficulty of transmitting electrons from the left terminal to the right, directly via the two current electrodes. This makes the crucial difference, i.e., positive definiteness, as compared to $R_{xx}$. Because $\tilde{R}_{xx}$ is directly associated with the current flowing through the entire sample, it must convert electric power to Joule heat and obey the first and second thermodynamic laws, which is then guaranteed by $\tilde{R}_{xx}>0$. On the other hand, $R_{xx}$ is fundamentally not subject to these stringent constraints since it is not directly associated with the complete current flow through the system.

\subsection{Symmetry relations}\label{App:symmetry}
Without much loss of generality, we take $\sigma$'s in the model Eq.~\eqref{eq:H_1band1} as pseudospins for certain orbitals. This simplifies the discussion below but is not fundamentally necessary since we do not need to specify the physical representation basis of $\sigma$. The relevant are mainly the mirror reflection $\cM_x$ along $x$-axis, spatial inversion $\cI$, and time reversal $\cT$. For the nonce, we consider a more generalized form of Eq.~\eqref{eq:H_1band1} by including a chirality parameter $\chi=\pm1$
\begin{equation}\label{eq:H_1band1_chi}
    H(\bm{k}) = \sum_{i=x,y,z}2 D_i(1-\cos{k_i})\sigma_0 + A(\chi\sin{k_x}\sigma_x + \sin{k_y}\sigma_y) +2M[(1-\cos k_\text{w}) - \sum_{i=x,y,z} (1-\cos k_i)]\sigma_z. 
\end{equation}
Given the symmetry properties below, we can exemplify a few representative relations. 
\begin{itemize}
\item The system itself is not invariant under $\cM_{x}$ and the magnetic field $\bB$ is reversed to $-\bB$, which actually satisfies 
\begin{equation}\label{eq:Mx}
    \cM_x H(\chi,\bB)\cM_x^{-1}=H(-\chi,-\bB).
\end{equation}
The transmission $1\rightarrow0$ is geometrically mapped to $2\rightarrow3$ by $\cM_x$. According to Eq.~\eqref{eq:Mx}, we are led to the relation $T_{01}(\chi,\bB)=T_{32}(-\chi,-\bB)$. Further using the Onsager reciprocity in the presence of both intrinsic and extrinsic $\cT$-breaking
\begin{equation}\label{eq:Onsager}
    T_{ij}(\chi,\bB)=T_{ji}(-\chi,-\bB)
\end{equation}
as per Eq.~\eqref{eq:T}, we arrive at 
\begin{equation}
    T_{01}(\bB)=T_{23}(\bB)
\end{equation}
for our system with $\chi=1$.

\item The system and also the magnetic field are invariant under $\cI$. The transmission $2\rightarrow0$ is geometrically mapped to $5\rightarrow3$ by $\cI$ and we are directly led to the relation 
\begin{equation}
    T_{02}(\bB)=T_{35}(\bB)
\end{equation}
since $\bB$ remains invariant under $\cI$. 

\item The system itself breaks $\cT$ and the magnetic field $\bB$ is reversed to $-\bB$, which actually satisfies 
\begin{equation}\label{eq:T}
    \cT H(\chi,\bB)\cT^{-1}=H(-\chi,-\bB).
\end{equation}
The transmission $2\rightarrow0$ is geometrically mapped to $4\rightarrow0$ and $\bB$ is reversed by the combination $\cI\cM_x$. Similar to i), we are eventually led to the relation 
\begin{equation}
    T_{02}(\bB)=T_{40}(\bB)
\end{equation}
again with the Onsager reciprocity Eq.~\eqref{eq:Onsager} used. 

\item As a digression, we also discuss an asymmetry relation here. We can consider the mirror reflection $\cM_y$ along $y$-axis. The system itself is not invariant under $\cM_{y}$ while the magnetic field $\bB$ is invariant, which actually satisfies 
\begin{equation}\label{eq:My}
    \cM_y H(A,\chi,\bB)\cM_y^{-1}=H(-A,-\chi,\bB).
\end{equation}
The transmission $1\rightarrow5$ is geometrically mapped to $5\rightarrow1$ by $\cM_y$. According to Eq.~\eqref{eq:My}, we are led to the relation $T_{51}(A,\chi,\bB)=T_{15}(-A,-\chi,\bB)$, i.e., one in general would \textit{not} always expect
\begin{equation}\label{eq:T51}
    T_{51}(\bB)= T_{15}(\bB)
\end{equation}
to hold. 
Another possibility is to consider the combination $\cM_x\cI$, which gives 
\begin{equation}\label{eq:MxI}
    \cM_x\cI H(A,\chi,\bB)\cI^{-1}\cM_x^{-1}=H(A,-\chi,-\bB).
\end{equation}
It also maps transmission $1\rightarrow5$ to $5\rightarrow1$ although in a way slightly different from $\cM_y$, where the difference is insignificant to our purpose. Then we are led to the relation $T_{51}(A,\chi,\bB)=T_{15}(A,-\chi,-\bB)$, which implies the possible breaking of Eq.~\eqref{eq:T51} again.

\end{itemize}
In this spirit, we have a whole set of symmetry relations for all $T$'s ($\bB$ dependence omitted):
\begin{equation}\label{eq:symmetry_diag}
    T_{3} = T_{0}, T_{2} = T_{1}, T_{4} = T_{1}, T_{5} = T_{1}, 
\end{equation}
for the diagonal and 
\begin{equation}\label{eq:symmetry_offdiag}
\begin{split}
  &T_{23} = T_{01}, 
  T_{34} = T_{01}, T_{50} = T_{01}, T_{13} = T_{02}, T_{35} = T_{02}, 
  T_{40} = T_{02}, T_{30} = T_{03}, T_{20} = T_{04}, T_{31} = T_{04}, 
  T_{53} = T_{04}, \\
  &T_{10} = T_{05}, T_{32} = T_{05}, 
  T_{43} = T_{05}, T_{45} = T_{12}, T_{25} = T_{14}, T_{41} = T_{14}, 
  T_{52} = T_{14}, T_{42} = T_{15}, T_{54} = T_{21}, T_{51} = T_{24}
\end{split}
\end{equation}
for the off-diagonal. Eq.~\eqref{eq:symmetry_diag} and Eq.~\eqref{eq:symmetry_offdiag} are not fully independent: the $T_{3}$ and $T_{4}$ relations in Eq.~\eqref{eq:symmetry_diag} are implied by Eq.~\eqref{eq:symmetry_offdiag} while the $T_{2}$ and $T_{5}$ relations in Eq.~\eqref{eq:symmetry_diag} are not.
In fact, the off-diagonal Eq.~\eqref{eq:symmetry_offdiag} suffices to enable Eq.~\eqref{eq:6T_Rxx}.

\subsection{Comparison with Dirac semimetal}\label{App:DSM}

\begin{figure}[hbt]
\includegraphics[width=11.5cm]{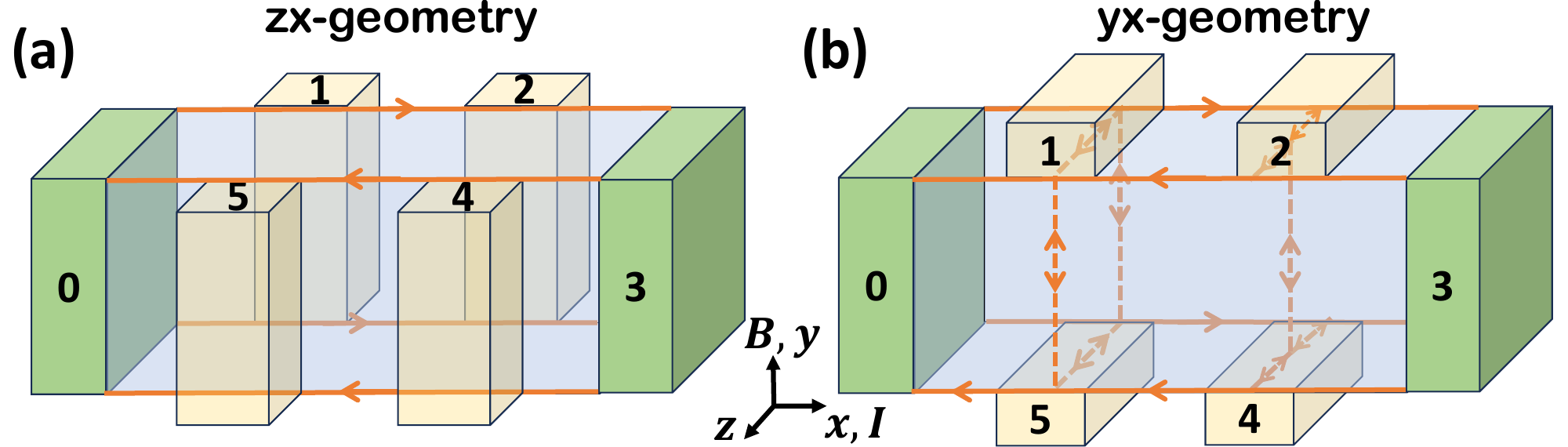}
\caption{Two Hall-bar configurations for a 3D DSM with the magnetic field (a) perpendicular and (b) parallel to the measurement plane, corresponding to Fig.~\ref{Fig:cartoon} for a WSM. The number of hinge states is doubled; two diagonal pairs exist along all four hinges, undermining the unique chiral nonlocal backscattering path in Fig.~\ref{Fig:cartoon}(b).
}\label{Fig:DSM_device}
\end{figure}

\begin{figure}[hbt]
\includegraphics[width=15.5cm]{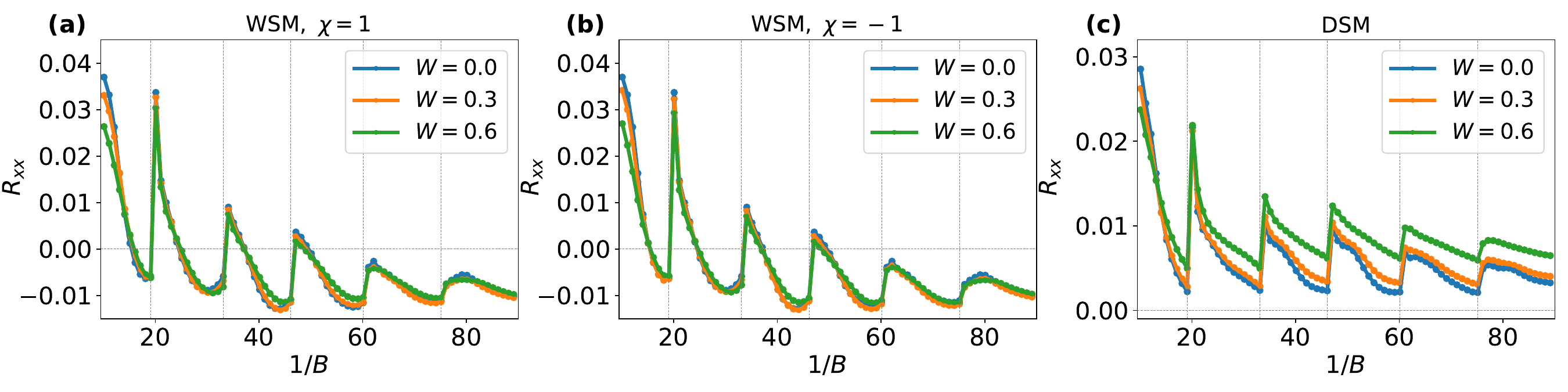}
\caption{Comparing the longitudinal resistance $R_{xx}$ measured in $yx$-geometry between WSM and DSM. (a) WSM with chirality $\chi=1$, (b) WSM with $\chi=-1$, (c) DSM. The vertical dashed lines mark the magnetic fields of $zx$-QHE plateau transitions from the measurement in Fig.~\ref{Fig:DSM_device}(a). System size is $28\times20\times20$ and three disorder strengths $W$ are shown. Other parameters are identical with Fig.~\ref{Fig:Ryx_Rxx}.
}\label{Fig:DSM_compare_Rxx}
\end{figure}

Because of the existence of similar hinge states and CLLs in Dirac semimetals (DSMs), as shown in Fig.~\ref{Fig:DSM_device}, it has long been deemed that the transport features related to the 3D QHE physics are mostly the same as those in WSMs, besides hinge modes doubled in number. Due to the degeneracy lifting of the two pairs of hinge states, the DSM often does not lead to even-integer QHE\cite{Zhang2018,Nishihaya2021}. Hence, there is no known way to distinguish WSM and DSM in terms of the 3D QHE. Surprisingly, the negative longitudinal resistance can exactly fill this gap.

The Hamiltonian of DSM can be composed by Eq.~\eqref{eq:H_1band1_chi} with two opposite chiralities, written as
\begin{equation}\label{eq:H_DSM}
    H_{\text{DSM}}(\bm{k}) = \sum_{i=x,y,z}2 D_i(1-\cos{k_i})\tau_0\sigma_0 + A(\sin{k_x}\tau_z\sigma_x + \sin{k_y}\tau_0\sigma_y) +2M[(1-\cos k_\text{w}) - \sum_{i=x,y,z} (1-\cos k_i)]\tau_0\sigma_z,
\end{equation}
where $\tau$-Pauli matrices represent the extra degrees of freedom apart from $\sigma$. For better comparison, the parameters are chosen the same as those of WSMs. As shown in Fig.~\ref{Fig:DSM_compare_Rxx}, the longitudinal resistances $R_{xx}$ for WSM show the same positive-negative oscillating behavior for both chiralities. In stark contrast, although $R_{xx}$ also exhibits an oscillatory behavior for DSM, its value always remains positive. 

Recalling the discussion in Sec.~\ref{App:6T}, one can understand the absence of the peculiar negative longitudinal resistance in DSM. The key to $R_{xx}<0$ in WSM lies in the anomalous transimission probability relations $T_{23}<T_{13},T_{51}<T_{15}$. They entirely rely on the very existence of the 3D nonlocal backscattering trajectory in Fig.~\ref{Fig:cartoon}(b), which is left-handed with respect to the positive $\hat{z}$-direction. However, in Fig.~\ref{Fig:DSM_device}(b), the uniqueness of such transporting path is diminished: right-moving carriers, for instance, can be transmitted back through both left-handed and right-handed trajectories, e.g., 0-1-5-0 and 0-5-1-0; also, there exist backscattering paths not making use of the CLLs. Therefore, the anomalous transmission probability cannot hold; the anomalous imbalance of chemical potentials $\mu_1<\mu_2$, due to the unconventional equilibration with source/drain carriers dictated by the unique 3D backscattering path, will be lost. On the other hand, these complications are also expected to affect the Hall quantum oscillation in $R_{yx}$ and the linear two-terminal magnetoresistance in $\tilde{R}_{xx}$. For instance, the former can vanish since the pair of time-reversal partners can contribute oppositely as the field is in-plane.
The latter will persist roughly because the existence of these backscattering paths suffices to modify the quantized behavior. The detailed behavior will be left for future detailed studies.

\section{Additional transport phenomena and experimental proposals}\label{App:ExpConfig}
\subsection{Recovery of quantized Hall signal and its sign change}

\begin{figure}[hbt]
\includegraphics[width=14.5cm]{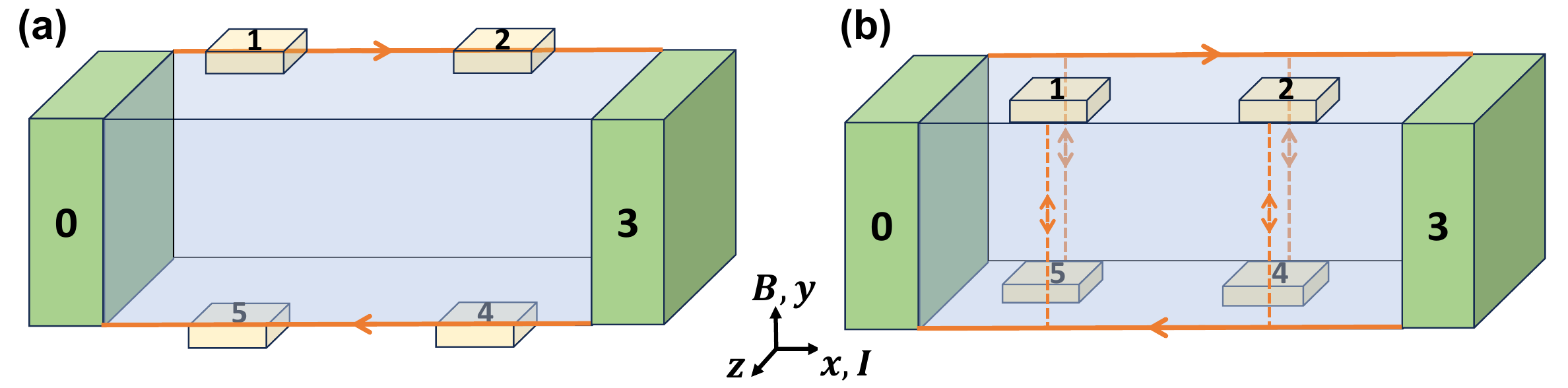}
\caption{The enlarged illustration of the two lower insets in Fig.~\ref{Fig:Rxy_recover_quantized}.
}\label{Fig:enlarged_insets}
\end{figure}

We give in Fig.~\ref{Fig:enlarged_insets} the enlarged illustration of the two lower insets in Fig.~\ref{Fig:Rxy_recover_quantized}, which recover the quantization of Hall resistance $R_{yx}$ in the $yx$-geometry and show a sign change when the lead arrangement is changed from (a) to (b). Disorders can affect such CLL-mediated connections and also generate inter-CLL scatterings.
The solid orange lines represent the hinge states. The dashed vertical orange lines represent the CLLs, via which the voltage leads connect to the hinge states in (b). Here, the Hall resistance, the two-terminal resistance, and the longitudinal resistance still follow the definition in Eq.~\eqref{eq:standard_R_definition}. In case (b), top leads-1,2 instead measure the voltage of bottom hinge states and hence of the drain, via vertical CLLs below these leads; bottom leads-4,5 are similarly connected to the top hinge states and hence the source. Thus, it still effectively detects the QHE but realizes a reversed lead-hinge connection.

\begin{figure}[hbt]
\includegraphics[width=14.5cm]{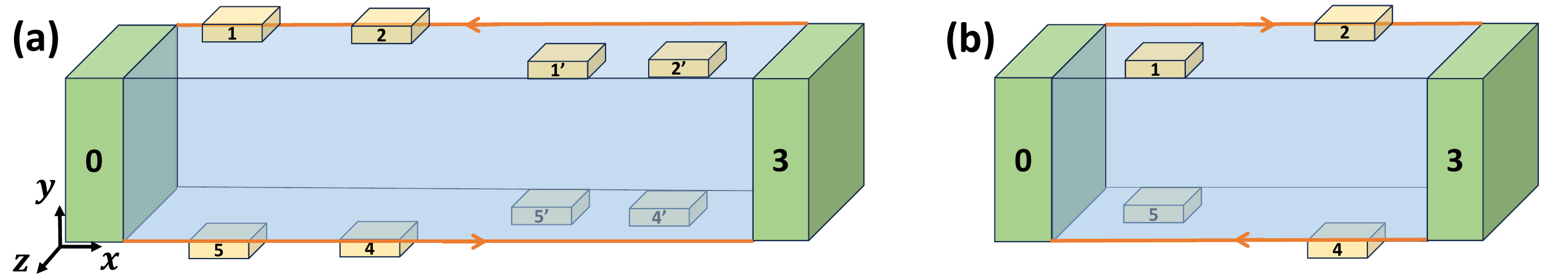}
\caption{(a) The combined device of Fig.~\ref{Fig:Rxy_recover_quantized}. (b) The simplified device with $yx$-geometry. 
}\label{Fig:exp-device}
\end{figure}

The features discussed in the main text, if experimentally observed, can be used as strong indications of the unique 3D quantum Hall mechanism. In particular, we further comment on the recovery of the quantized Hall resistance as a possible smoking-gun evidence in experiments, where the Hall resistance undergoes a \textit{sign change} as the lead configuration changes, as exemplified in Fig.~\ref{Fig:Rxy_recover_quantized}. For experimental convenience, one can prepare a device that combines the two devices in Fig.~\ref{Fig:Rxy_recover_quantized}, as shown in Fig.~\ref{Fig:exp-device}(a). Alternatively, one can use another six-terminal device in Fig.~\ref{Fig:exp-device}(b), exploiting the essentials of the 3D CLL-assisted mechanism. 
In the $yx$-geometry, this device will show sign changes and give 
\begin{equation}
    R_{15}=-R_{24}=R_{12}=-R_{54}=-\frac{1}{n}\frac{h}{e^2},
\end{equation}
which would be drastically different if it were in the $zx$-geometry in Fig.~\ref{Fig:cartoon}(a)
\begin{equation}
    R_{15}^{(zx)}=R_{24}^{(zx)}=\frac{1}{n}\frac{h}{e^2},\quad R_{12}^{(zx)}=R_{54}^{(zx)}=0.
\end{equation}

\subsection{Hybrid lead configuration with nonquantization and negative resistance}\label{App:hybrid}
\begin{figure}[hbt]
\includegraphics[width=12.5cm]{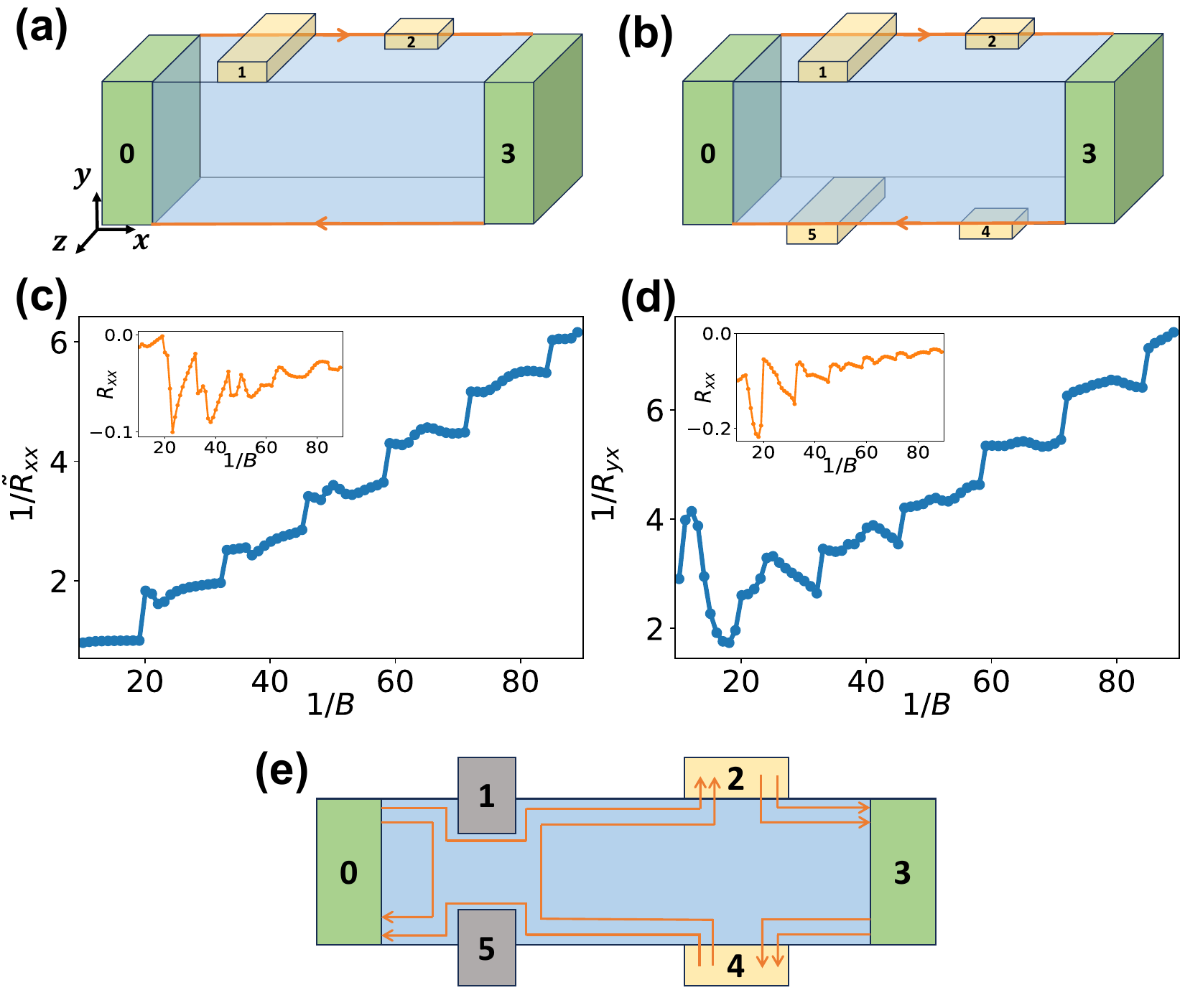}
\caption{Four-terminal (a) and six-terminal (b) device with hybrid lead configurations in $yx$-geometry. (c) The inverse two terminal resistance $1/\tilde{R}_{xx}=1/R_{03}$ of (a). (d) The inverse of the Hall resistance $1/R_{yx}=1/R_{24}$ of (b). Insets are the corresponding longitudinal resistance $R_{xx}=R_{12}$. (e) 2D Hall bar with the grey split gate constriction at positions 1 and 5; solid lines indicate 2D QHE edge channels, in which one channel is backscattered as an example.
}\label{Fig:hybrid-device}
\end{figure}

As mentioned in the main text, we can also set up devices with hybrid lead configurations, which, in contrast to the conventional QHE, fail to give Hall quantization but show robust negative longitudinal resistance as seen in Fig.~\ref{Fig:hybrid-device}(c,d).

For the four-terminal device as shown in Fig.~\ref{Fig:hybrid-device}(a), the transmission probability from lead-3 to lead-2 is negligibly small, since it is the antichiral transmission with a narrow lead-2. The narrowness of lead-2 prevents it from significantly taking advantage of the vertical bulk CLLs, making $T_{23}$ very similar to the antichiral transmission in 2D QHE. This differs from the wide lead-1, which, due to partially the 3D backscattering path incident from the right, can readily take advantage of the bottom left-moving chiral hinge states and the vertical CLLs, making $T_{13}$ nonnegligible although seemingly antichiral. Therefore, from Eq.~(\ref{eq:sgn_4T_Rxx}), $\sgn(R_{xx})=\sgn(T_{10}T_{23}-T_{20}T_{13}) \approx \sgn(-T_{20}T_{13}) < 0$. This enables the negative $R_{xx}$ in Fig.~\ref{Fig:hybrid-device}(c).

For the six-terminal device in Fig.~\ref{Fig:hybrid-device}(b), the top leads-1,2 have the same configuration as that of the four-terminal one in Fig.~\ref{Fig:hybrid-device}(a). Due to leads-1,5, we find the 3D backscattering trajectory incident from the right, i.e., $4\rightarrow5\rightarrow1\rightarrow2$, at work but not the reverse. Wide lead-5 thus brings electrons from the drain to lead-1 through hinge modes, interface conducting channels and bulk CLLs; the chemical potential $\mu_1$ of lead-1 will be a definite average between the source and drain chemical potentials $\mu_s$ and $\mu_d$, i.e., $\mu_1<\mu_s$. As is physically clear, fewer $\mu_d$-carrying electrons can further reach lead-2 beyond lead-1; narrow lead-4 exerts a weaker effect upon lead-2 and one would thus expect $\mu_2$ not far away from $\mu_s$ carried by the top hinge mode, i.e., $\mu_2\lesssim \mu_s$. Therefore, the system keeps the negative resistance $R_{xx}=R_{12}$ robust since $\mu_1<\mu_2$ generally holds, as seen in Fig.~\ref{Fig:hybrid-device}(d). 
From the perspective of Fig.~\ref{Fig:exp-device}(b), which has a negative quantized $R_{12}$, the present device can be thought of as lead-1 and lead-5 gradually expanding to their full width. Meanwhile, although the effects of mixing the electrons from the source and drain via bulk CLLs set in and break the perfect quantization, it still retains the negativity of $R_{12}$.

As mentioned in the main text, the setup in Fig.~\ref{Fig:hybrid-device}(b) is very similar to a 2D QHE Hall-bar system in the $xy$-plane illustrated in Fig.~\ref{Fig:hybrid-device}(e). The classical result in 2D states that Hall quantization for $R_{24}$ still remains despite the backscattering due to split gate constriction in the left region.
The unconventional breaking of Hall quantization in our setup can be understood as follows. 
The foregoing 3D backscattering trajectory  $4\rightarrow5\rightarrow1\rightarrow2$ means an asymmetry $T_{42}< T_{24}$ different from what we have seen in Sec.~\ref{App:6T}, which affects the equilibration in lead-2 and lead-4: bottom-hinge electrons with drain chemical potential sneak into lead-2 together with normal source electrons from the left while lead-4 basically only sees electrons from the drain. Thus, the premise in 2D QHE that the top and bottom leads-2,4 are equally connected to the hinge modes is no longer true, hence undermining the quantization as seen in Fig.~\ref{Fig:hybrid-device}(d).

\section{Transport features versus topological numbers}\label{App:transportvstopo}
\subsection{Other transport measurement planes}\label{App:othertransport}

\begin{figure}[hbt]
\includegraphics[width=14.8cm]{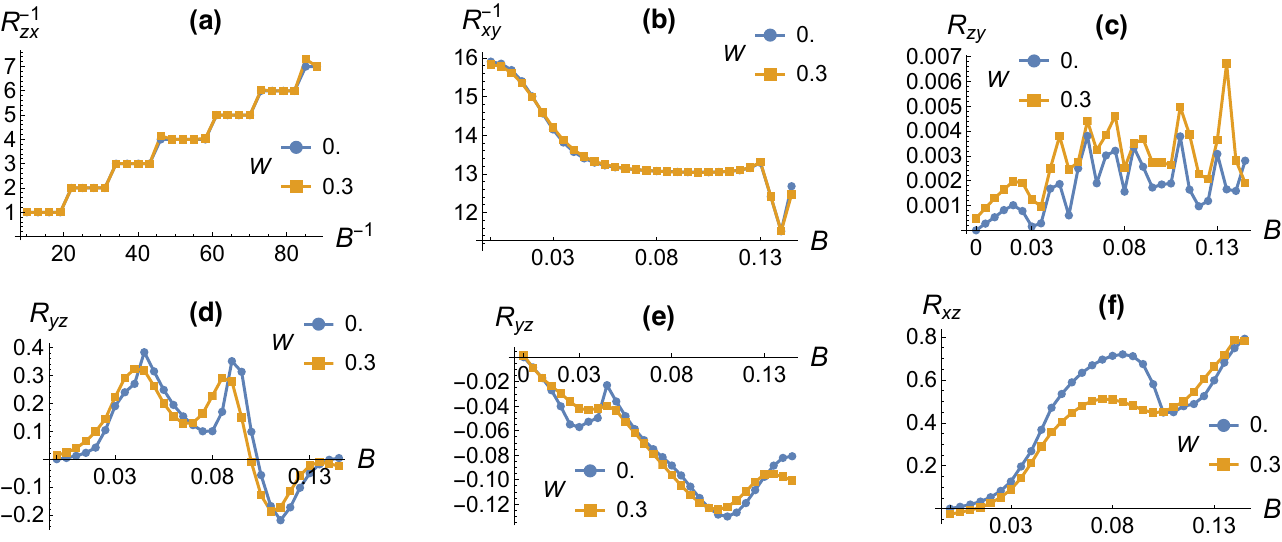}
\caption{Essential Hall-channel transport features in geometries other than the $yx$-geometry discussed in the main text. Top row: (a) $R_{zx}$ displays the 3D QHE; (b) $R_{xy}$ shows the AHE of WSM; (c) $R_{zy}$ is vanishingly small compared to all other Hall signals and due to finite size effects; they match the predictions from 3D topological Chern number calculations in Table~\ref{tab:3DChern}.
Bottom row: (d) $R_{yz}$ shows finite Hall signals and (e) the Hall signal varies and reduces with a larger size of voltage leads; (f) $R_{xz}$ does not show any Hall quantization; they, together with $R_{yx}$ in the main text, do not match those predictions as indicated also in Table~\ref{tab:3DChern}.
}\label{Fig:othergeometry}
\end{figure}

\begin{itemize}
    \item $zx$-geometry

This geometry gives the perfect 3D WSM QHE as shown in Fig.~\ref{Fig:othergeometry}(a) and as predicted by the 3D Chern number in Fig.~\ref{Fig:CxCyCz}(a).

    \item $xy$-geometry

In WSMs that break the time-reversal symmetry $\cT$, an intrinsic AHE will appear in the measurement plane orthogonal to the WP alignment, i.e., $xy$-plane in our system. This is due to the quantum anomalous Hall effect (QAHE) from each momentum $k_z$-slice 2D system with circulating edge channels responsible for the Fermi arcs. This effect is accumulated between the WP pair and the circulating states are distributed near the $yz$- and $xz$-surfaces in the real space. The corresponding Hall conductivity reads 
\begin{equation}
    \sigma_{xy}=\frac{1}{L_z}\sum_{k_z}\sigma_{xy}^\mathrm{2D}(k_z)=\frac{2k_w}{2\pi}\frac{e^2}{h}
\end{equation}
as per Eq.~\eqref{eq:H_1band1}, which equals $\frac{e^2}{2h}$ since we set $k_w=\pi/2$ and is also corroborated by Fig.~\ref{Fig:CxCyCz}(a). 
As elaborated in the main text, this does not make any meaningful prediction for the $yx$-geometry. This is because the relevant $xz$-surfaces are orthogonal to $B\hat{y}$; hence, the arc states are magnetically quantized into arc LLs, whose edge states become the hinge states.
On the other hand, the $xy$-geometry is simpler since the relevant $yz$-surfaces are parallel to $B\hat{y}$ and thus not readily affected by applying an in-plane magnetic field, besides certain less important band distortions.
Then the inverse Hall resistance $R_{xy}^{-1}$ 
will approximately be given by $\lfloor L_z\sigma_{xy}\rfloor$ with $\lfloor\cdots\rfloor$ the conventional integral floor function. This is indeed seen in Fig.~\ref{Fig:othergeometry}(b), where $R_{xy}^{-1}$ at small $B$ is very close to the theoretical value $16$ and remains around the same order of magnitude even at large $B$.

    \item $zy$-geometry

As shown in Fig.~\ref{Fig:othergeometry}(c), $R_{zy}(W=0)$ without disorder remains very close to zero, bearing the smallest order of magnitude in all Hall resistances in this system; the tiny signals can be verified to originate from finite size effects in the lattice simulation. Physically, the CLLs directly contribute to the current transport. Also, the Hall voltage is measured on the front and back $xy$-surfaces parallel to $B\hat{y}$, where neither nontrivial QAHE intrinsic edge circulating channels nor LLs can exist. This therefore matches the Chern number $\cC_x=0$ found in Fig.~\ref{Fig:CxCyCz}(a), as shown in Table~\ref{tab:3DChern}.

    \item $yz$-geometry

As shown in Fig.~\ref{Fig:othergeometry}(d), $R_{yz}$ is typically finite. Different from the $zy$-geometry, now the CLLs do not contribute to current transport and the Hall voltage is measured on the top and bottom $xz$-surfaces orthogonal to $B\hat{y}$. The nontrivial QAHE intrinsic edge circulating channels are present and magnetically quantized to LLs on these surfaces, but the hinge modes are orthogonal and hence do not contribute to the current transport along $z$-direction. Or more explicitly, if the current leads cover the $xy$-surfaces, there are no hinge modes. The presence of those nontrivial $xz$-surfaces is possible to give unquantized field-induced Berry curvatures near the top and bottom edges of each $yz$-planes, dependent on the details of how the band structure is modified by the field, which are picked up by the Hall signal. Note that such a nonuniversal contribution due to the existence of surface is absent in the $zy$-geometry with voltage leads on the trivial surfaces; also, it is not picked up by the $\cC_x$ calculation below since such a surface effect with no topological correspondence is not included, no matter whether periodic or open boundary condition is used. In addition, 
larger voltage leads are expected to be subject to a stronger short-circuiting effect and hence smaller Hall voltage. This is indeed seen in Fig.~\ref{Fig:othergeometry}(e), where the voltage leads are enlarged about five times in the area, which again signifies the nonuniversal nature of this Hall signal. Such physical differences clearly lead to $R_{yz}\neq R_{zy}$ and thus are different from the expectation from $\cC_x=0$, as shown in Table~\ref{tab:3DChern}.

    \item $xz$-geometry

As shown in Fig.~\ref{Fig:othergeometry}(f), $R_{xz}$ is obviously different from the expectation of 3D QHE like Fig.~\ref{Fig:othergeometry}(a) or the quantized $\cC_y$ as shown in Table~\ref{tab:3DChern}. The core factor is the nontrivial side $yz$-surfaces, which are inevitably metallic and thus destroy the flat LL gaps. Such metallicity is, however, indispensable to the $zx$-geometry 3D QHE since it gives rise to the curved Fermi arcs linked to the Weyl orbit, i.e., the absence of QHE in the $xz$-geometry is an intrinsic property.
\end{itemize}

\subsection{The 3D topological characterization}\label{App:Chern}

\begin{figure}[hbt]
\includegraphics[width=11.4cm]{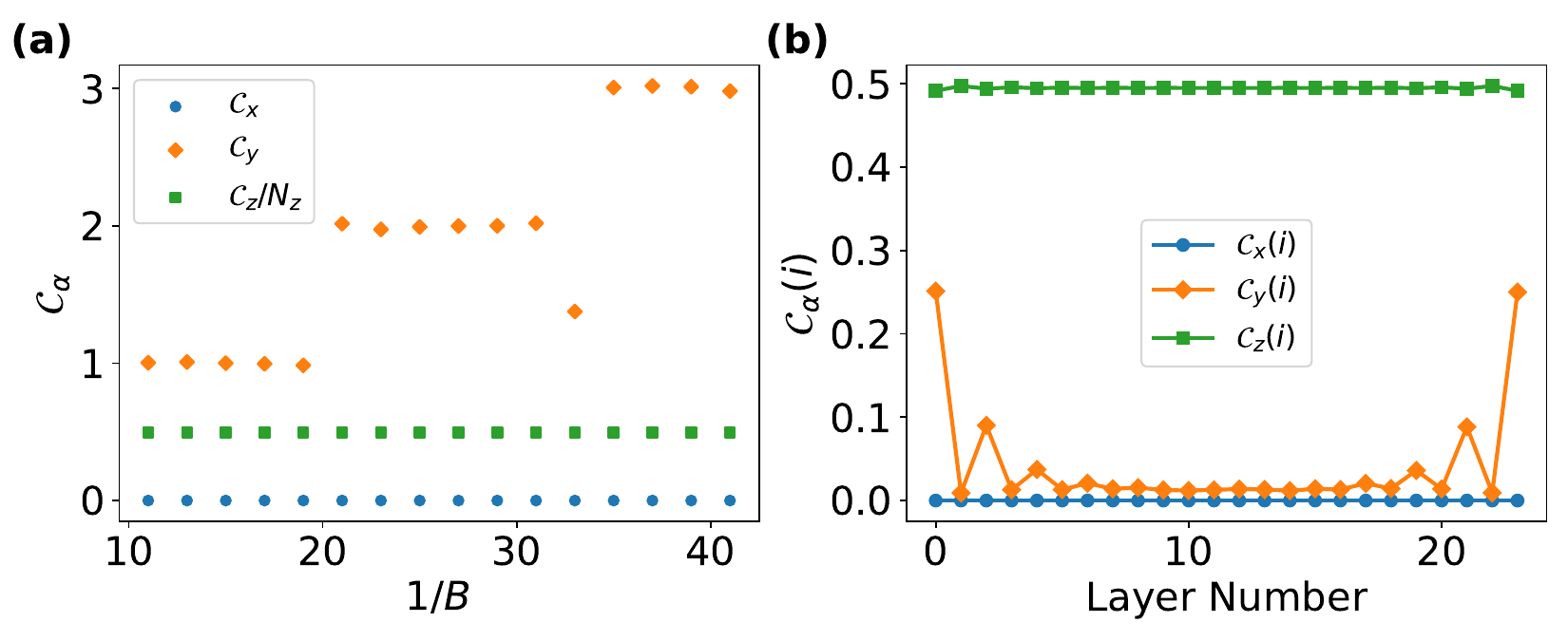}
\caption{(a) The 3D Chern number $\cC_\alpha$ in three directions ($\alpha=x,y,z$). (b) The layered-resolved Chern number $\cC_\alpha(i)$ in three directions with $1/B=15$. For the calculations of $\cC_x$ and $\cC_z$, the system size is chosen to be $24\times 24 \times 24$ using Eq.~(\ref{eq:non-comm-chern}); while for $\cC_y$, we use Eq.~(\ref{eq:non-comm-chern2}) with size $300\times 24 \times 300$, $M=6000$ and $R=40$.
}\label{Fig:CxCyCz}
\end{figure}

In this section, we calculate the 3D Chern numbers in three orthogonal directions. The result is shown in Fig.~\ref{Fig:CxCyCz}. Since the magnetic field is along $\hat{y}$, when calculating the 3D Chern number $\cC_x$ along $x$-direction, we choose the gauge $\bA = (0, 0, -Bx)$ such that any layer in the $yz$-plane can use a periodic boundary condition. For the same reason, we choose the gauge $\bA = (Bz, 0, 0)$ for the 3D Chern number $\cC_z$ along $z$-direction. The layered Chern number is calculated by the real-space non-commutative Chern number method\cite{Prodan2009, Prodan2011}
\begin{equation}
    \cC_z(z_i)=2\pi \ii\mathrm{Tr}_{z_i}\hat{P}[-i[\hat{x},\hat{P}],-i[\hat{y}, \hat{P}]].\label{eq:non-comm-chern} 
\end{equation}
Here, the periodic boundary condition is taken in both $x$ and $y$ directions. $\hat{P}=\theta(E_f-H)$ is the projection operator of the occupied states of the Hamiltonian, and $\mathrm{Tr}_{z_i}$ is the trace over the states in the $z_i$-th layer. The total 3D Chern number $\cC_{z}$ is obtained by summing up all the contributions from each layer, $\cC_z = \sum_{z_i} \cC_z(z_i)$. Similar formulae also apply for $\cC_x(x_i)$ and $\cC_x$. One might wonder whether the periodic boundary condition is enough to capture the hinge modes. For instance, in the calculation of $\cC_z$, no $z_i$-layer explicitly bears any part of the hinge mode. However, it is important to understand that, in each $z_i$-layer, it is the conventional bulk-boundary correspondence at work that captures the boundary states from purely bulk calculation; any second-order topological hinge mode as a whole is eventually captured by further summing up all the $z_i$-layers; in other words, the second-order topology is reduced to first-order within each $z_i$-layer at the price of an extra summation over $z_i$-layers. Also, this is consistent with the following fact to be mentioned shortly below: even open boundary conditions do not directly incorporate the boundary, because sites near the edges must be excluded. Therefore, one expects identical results from both types of boundary conditions, if the condition is possible, which is also confirmed in our explicit calculation.

\begin{figure}[hbt]
\includegraphics[width=16.4cm]{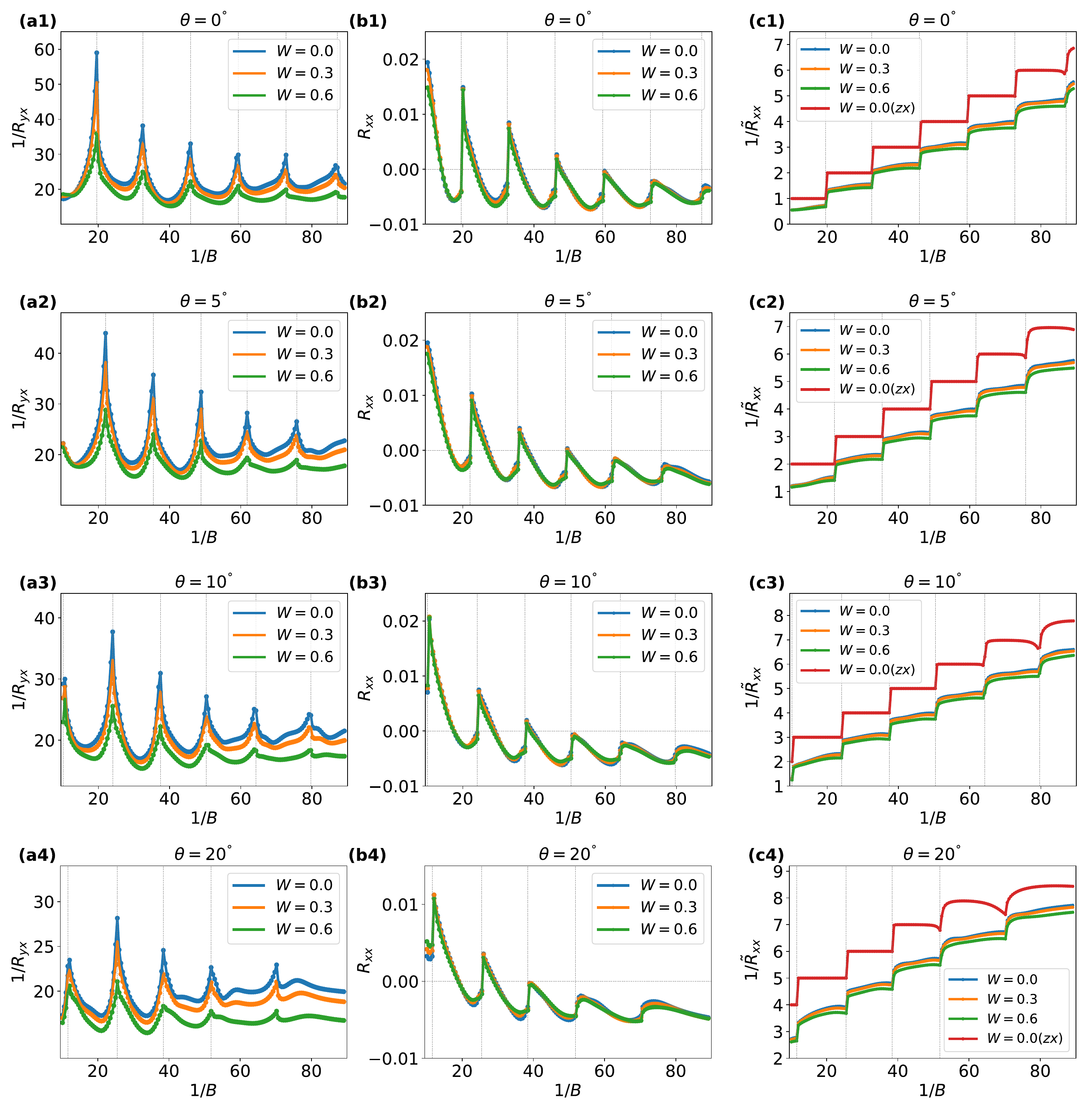}
\caption{Transport properties in the $yx$-geometry with rotated magnetic field $\bm{B} = B(0, \cos\theta, \sin\theta)$. (a1)-(c1) $\theta=0^\degree$, (a2)-(c2) $\theta=5^\degree$, (a3)-(c3) $\theta=10^\degree$, (a4)-(c4) $\theta=20^\degree$. The vertical dashed lines mark the magnetic fields of $zx$-QHE plateau transitions. System size is $24\times24\times24$ and three disorder strengths $W$ are shown. Other parameters are identical to Fig.~\ref{Fig:Ryx_Rxx}.
}\label{Fig:rotate_B}
\end{figure}

When calculating the 3D Chern number $\cC_y$ along $y$-direction, the vector potential $\bA$ inevitably breaks the translational symmetry along either $x$ or $z$ direction, thus the periodic boundary condition cannot be applied in general. One may apply the periodic boundary condition for the fine-tuned rational magnetic flux; however, the associated large system size beyond exact diagonalization when $B$ is small will make the projection operator $P$ in Eq.~\eqref{eq:non-comm-chern} difficult to obtain. To calculate $\cC_y$, we use the kernel polynomial method\cite{Weise2006, Varjas2020}, which allows us to calculate system size much larger than the typical size of exact diagonalization. The layered Chern number can be calculated using the expression given in\cite{Bianco2011, Varjas2020}
\begin{equation}\label{eq:non-comm-chern2} 
    \cC_y(y_i) = -2\pi \ii \mathrm{Tr}_{y_i}' [\hat{P}\hat{z}\hat{P}, \hat{P}\hat{x} \hat{P}].
\end{equation}
Here, we take the open boundary condition. 
$\mathrm{Tr}_{y_i}'$ is the trace over the subsystem of layer $y_i$, excluding the sites near the edges to avoid vanishing results\cite{Bianco2011}. The projection operator $\hat{P}$ on a state $\ket{v}$ is calculated using the kernel polynomial method. Up to the order $M$, it can be expressed as
\begin{equation}
    \hat{P}(E_f, H) \ket{v} = \sum_{m=0}^{M} g_m \mu_m(\epsilon_f) \ket{v_m},
\end{equation}
where $\mu_m(\epsilon_f)$ is the $m$-th moments of $\hat{P}$ and takes the form 
\begin{equation}
    \mu_m(\epsilon_f)=\begin{cases}
        1-\frac{\arccos (\epsilon_f)}{\pi} &\quad m=0\\
        \frac{-2\sin(m\arccos(\epsilon_f))}{m\pi} &\quad m\neq 0
    \end{cases}.
\end{equation}
Here, $\epsilon_f$ is the rescaled Fermi energy lying in the interval $[-1,1]$ as
\begin{equation}
    \epsilon_f = \frac{2}{E_+-E_-} (E_f-\frac{E_+ + E_-}{2})
\end{equation}
with $E_+$ and $E_-$ the eigenenergy maximum and minimum of $H$.
The state $\ket{v_m}$ is recursively defined as
\begin{equation}
    \ket{v_0}=\ket{v}, \quad \ket{v_1}=\tilde{H}\ket{v_0} \quad ,\ket{v_{m+1}}=2\tilde{H}\ket{v_m}-\ket{v_{m-1}}
\end{equation}
with $\tilde{H}$ the rescaled Hamiltonian defined similarly as $\epsilon_f$.
The Jackson kernel $g_m$ takes the value
\begin{equation}
    g_m = \frac{(M-m+1)\cos(\frac{\pi m}{M+1})+\sin (\frac{\pi m}{M+1})\cot (\frac{\pi}{M+1})}{M+1}.
\end{equation}
To evaluate the trace, the stochastic trace approximation is used in the calculation\cite{Varjas2020}
\begin{equation}
    \mathrm{Tr}_S(\hat{O}) = \frac{1}{R|S|} \sum_{i=1}^R\bra{r_i}\hat{O}\ket{r_i},
\end{equation}
where $\ket{r_i}$ represents $R$ random-phase vectors localized in the region $S$ with area $|S|$.

\section{Transport properties with rotated magnetic field or sample}\label{App:rotated}
Previous calculations are based on the minimal model with the WP pair alignment normal to both $\bB$ and the current direction. In this section, we respectively rotate the magnetic field and the sample (equivalent to rotating the current). Since the crucial 3D nonlocal path is an intrinsic quantum channel of this WSM system under a magnetic field, we show that, in the rotated general cases, the proposed phenomena can still exist.

\subsection{Rotating the magnetic field}

The transport properties with a rotated magnetic field $\bm{B} = B(0, \cos\theta, \sin\theta)$ have been studied in Ref.~\cite{Li2020a} in the conventional $zx$-geometry, where quantized Hall plateaus have also been observed. In this section, we present the results of transport properties in the $yx$-geometry. \mycomment{To reduce the computational complexity, we choose the system size to be $24\times24\times24$.}

As shown in Fig.~\ref{Fig:rotate_B}, systems with a rotated magnetic field still possess all the same essential features as the case with a normal magnetic field $\bB=B\hat{y}$, which is as expected. The $zx$-QHE plateaus persist when $\bB$ is rotated, although the exact plateau values may vary with $\theta$ as shown in Fig.~\ref{Fig:rotate_B}(c1)-(c3), which is due to the varying number of contributing hinge channels as discussed in Ref.~\cite{Li2020a}. Likewise, since the chiral hinge states and CLLs are largely intact, the 3D backscattering paths can still form in the very same way and hence the similar phenomena in this case.

\subsection{Rotating the sample}
\begin{figure}[h!]
\includegraphics[width=7.6cm]{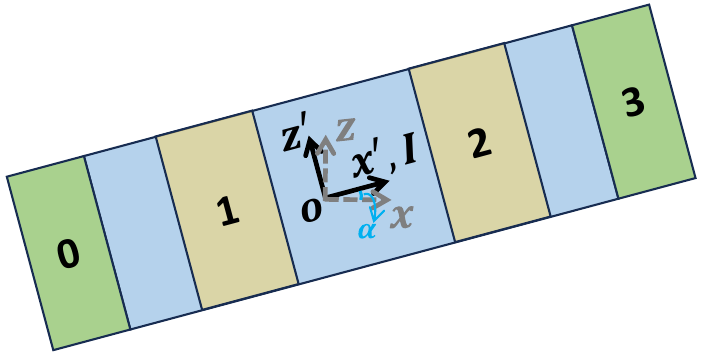}
\caption{Schematic of the device in the $yx$-geometry, where the Hall-bar system is rotated by an angle $\alpha$ around $y$-axis (normal to the page). Voltage lead-4 and lead-5 (not shown since covered by the sample) are beneath lead-2 and lead-1, respectively. The underlying lattice is unrotated (marked as $oxz$-coordinates); its $x$-direction has an angle $\alpha$ with respect to the current direction (rotated $x'$-direction). The whole system is projected onto the $xz$-plane for better illustration.
}\label{Fig:rotate_sys}
\end{figure}

\begin{figure}[hbt]
\includegraphics[width=16.4cm]{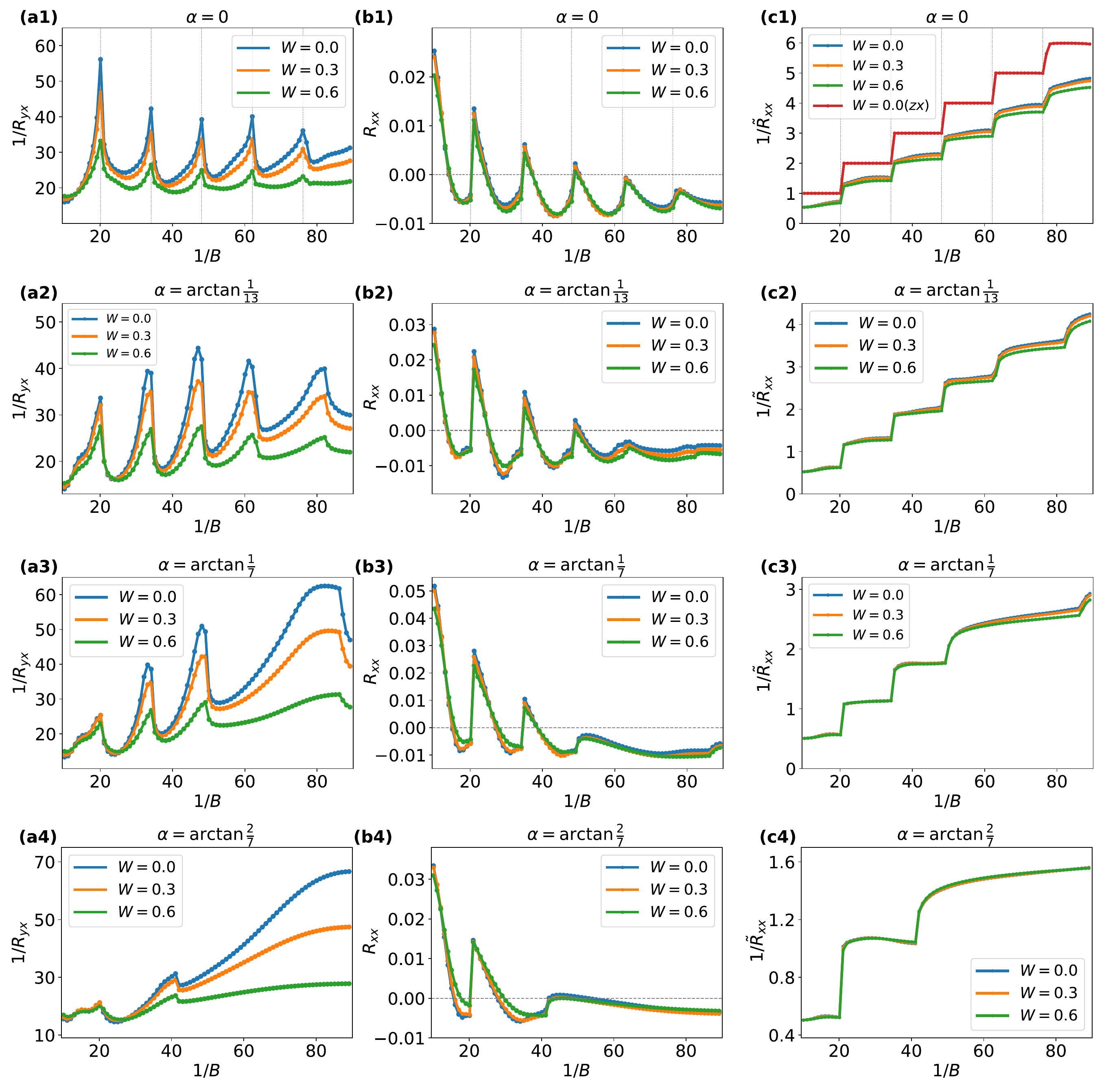}
\caption{Transport properties in the $yx$-geometry with the sample rotated by an angle $\alpha$ around $y$-axis. (a1)-(c1) $\alpha=0$, (a2)-(c2) $\alpha=\arctan(1/13)$, (a3)-(c3) $\alpha=\arctan(1/7)$, (a4)-(c4) $\alpha=\arctan(2/7)$. The vertical dashed lines in (a1)-(c1) mark the magnetic fields of $zx$-QHE plateau transitions; the horizontal dashed lines in (b1)-(b4) mark the zero resistance. System size is around $24\times15\times24$ and three disorder strengths $W$ are shown. Other parameters are identical to Fig.~\ref{Fig:Ryx_Rxx}.
}\label{Fig:rotate_sys_data}
\end{figure}

In this part, we present the transport results for the Hall-bar systems in the $yx$-geometry rotated around $y$-axis, as shown in Fig.~\ref{Fig:rotate_sys}, where the Hamiltonian is held fixed. Ref.~\cite{Zhang2022} focused on the $zx$-geometry under the same rotation and found that, when the current direction deviates from being normal to the WP pair alignment, the perfect Hall quantization will be lost and even no sign of QHE when they are parallel. This is due to mixing with the side-surface (normal to $x$-axis) metallic states, which are inevitable as they accompany the curved Fermi arc responsible for the 3D QHE. 
Intuitively, the features shown in Fig.~\ref{Fig:Ryx_Rxx} exactly avoid the Hall quantization, which should not be too sensitive to the mixing with the metallic states. This is in comparison to those conventional quantization features in $zx$-geometry, whose sole existence crucially depends on the absence of the metallic states. On the other hand, the metallic states will also bring their characteristics into the overall transport properties, which may show some competition with the properties inherited from the 3D nonlocal channels.

In Fig.~\ref{Fig:rotate_sys_data}, we show the results of the calculation where the rotation angle $\alpha$ is chosen to be \textit{commensurate} with the underlying lattice, i.e., it is of the form $\tan\alpha=\frac{m}{n}$ with $m,n\in\mathds{N}$; otherwise, one would not be able to maintain a complete and finite lattice system. Note that this is the only tractable way to simulate an oblique transport, although it is already computationally heavy.
The oscillatory profile of the in-plane Hall resistance $R_{yx}$ can be clearly identified in Fig.~\ref{Fig:rotate_sys_data}(a). In addition, it also shows a deviation from or reversal of the overall negative magnetoresistance, which is due to a competition between the metallic states and the pure effect of the proposed 3D-enabled nonlocal paths. These metallic states will contribute differently from the pure effect and hybridize with the other conducting channels. 
When the rotation angle is large, the portion of metallic states is large, resulting in an overall positive Hall magnetoresistance as shown in Fig.~\ref{Fig:rotate_sys_data}(a4). 
From Fig.~\ref{Fig:rotate_sys_data}(b), we find that the most peculiar feature, the negative longitudinal resistance $R_{xx}$, still holds well for the rotated systems, although the oscillation cycles are fewer at larger rotation angles. The two-terminal resistance $\tilde{R}_{xx}$ also exhibits the unquantized step-like profile with an upward slope within steps. The steps in the region of small magnetic fields might be missing for larger rotation angles, as in Fig.~\ref{Fig:rotate_sys_data}(c4), which is similar to the fewer oscillation cycles in both $R_{yx}$ and $R_{xx}$. 

\begin{figure}[hbt]
\includegraphics[width=7.cm]{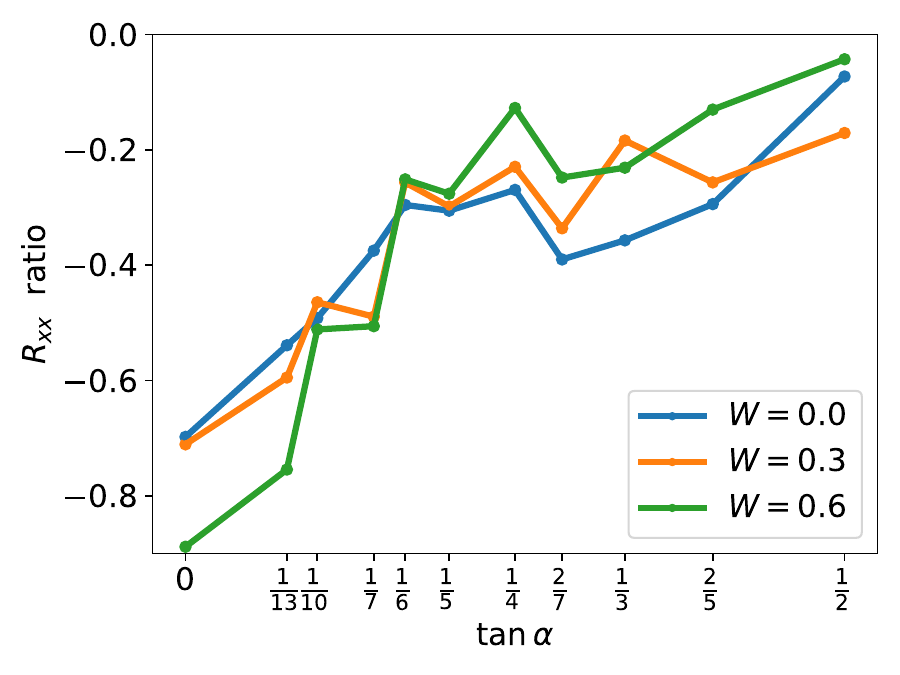}
\caption{Ratio of the average negative resistance to the average positive resistance of $R_{xx}$ in the $yx$-geometry with the sample rotated by an angle $\alpha$ around $y$-axis. System size is around $24\times15\times24$ and three disorder strengths $W$ are shown. Other parameters are identical to Fig.~\ref{Fig:Ryx_Rxx}.
}\label{Fig:maxRxx}
\end{figure}

The iconic feature of negative longitudinal resistance may be regarded as a fragile property due to its peculiar appearance. However, it turns out to be not only exceptionally robust to disorders, as shown in the main text, but also rather stable against the rotation discussed here. We present in Fig.~\ref{Fig:maxRxx} the ratio of the average negative $R_{xx}$ to the average positive $R_{xx}$ with respect to the rotation angle $\alpha$, for the same typical observation window of magnetic field as in Fig.~\ref{Fig:Ryx_Rxx}(b). The nonmonotonic behavior in Fig.~\ref{Fig:maxRxx} is largely due to the calculation method here, i.e., the lattice-commensurate rotation angle, which specifies a particular type of lattice geometry and hence introduces nonuniversal structures in signals. On the other hand, clearly, negative $R_{xx}$ persists even beyond $\alpha=\arctan\frac{1}{2}$, which promises plenty of room for experimental observation.

More importantly, one should keep in mind that the lattice size in these rotated calculations is chosen to be as small as $24\times 15\times 24$, merely a quarter of the sample volume used in the main text. However, calculations indicate that a larger system size can maintain all those features better. This is also related to the nonmonotonicity in Fig.~\ref{Fig:maxRxx}. However, inherent to the wavefunction scattering approach we adopted, this is unfortunately constrained by the exceedingly resource-demanding computation for the rotated sample case, much more challenging than the unrotated case with the underlying lattice and transport geometry aligned. Therefore, it is physically understood that the new features of our focus will persist much better than Fig.~\ref{Fig:rotate_sys_data} and Fig.~\ref{Fig:maxRxx} in any real system in the thermodynamic limit, although they already display the stability clearly.

\end{document}